\documentclass[pra,twocolumn,twoside,showpacs]{revtex4}

\usepackage{graphicx}
\usepackage{amsmath,amssymb}
\usepackage{ae}

\newcommand{\eq}[1]{\begin{equation}#1\end{equation}}
\newcommand{\eqmulti}[1]{\begin{equation}\begin{split}#1\end{split}\end{equation}}

\newcommand{\ket}[1]{\ensuremath{\,|{#1}\rangle}}

\newcommand{\matrixe}[3]{\ensuremath{\langle{#1}|\,{#2}\,|{#3}\rangle}}

\newcommand{\op}[1]{\ensuremath{\hat{\mathrm{#1}}}}
\newcommand{\adj}[1]{\ensuremath{{{#1}}^{\dag}}}
\newcommand{\conj}[1]{\ensuremath{{{#1}}^{\star}}}

\newcommand{\aO}{\ensuremath{\op{a}}}
\newcommand{\aaO}{\ensuremath{\adj{\op{a}}}}
\newcommand{\cO}{\ensuremath{\op{c}}}
\newcommand{\ccO}{\ensuremath{\adj{\op{c}}}}
\newcommand{\nO}{\ensuremath{\op{n}}}

\newcommand{\xO}{\ensuremath{\op{x}}}

\newcommand{\AO}{\ensuremath{\op{A}}}
\newcommand{\AAO}{\ensuremath{\adj{\op{A}}}}
\newcommand{\HO}{\ensuremath{\op{H}}}
\newcommand{\JO}{\ensuremath{\op{J}}}
\newcommand{\TO}{\ensuremath{\op{T}}}
\newcommand{\UO}{\ensuremath{\op{U}}}

\newcommand{\deltaO}{\ensuremath{\op{\delta}}}

\newcommand{\vV}{\ensuremath{\vec{v}}}
\newcommand{\xV}{\ensuremath{\vec{x}}}
\newcommand{\yV}{\ensuremath{\vec{y}}}

\newcommand{\nablaV}{\ensuremath{\vec{\nabla}}}

\newcommand{\IC}{\ensuremath{\mathcal{I}}}
\newcommand{\VC}{\ensuremath{\mathcal{V}}}

\newcommand{\ii}{\ensuremath{\mathrm{i}}}
\newcommand{\ee}{\ensuremath{\mathrm{e}}}

\newcommand{\SF}{\ensuremath{\textrm{s}}}
\newcommand{\cond}{\ensuremath{\textrm{c}}}
\newcommand{\gap}{\ensuremath{\textrm{gap}}}

\begin{document}

\title{Phase Diagram of Bosonic Atoms in Two-Color Superlattices}

\author{Robert Roth}
\affiliation{Institut f\"ur Kernphysik, Technische Universit\"at Darmstadt,
64289 Darmstadt, Germany}

\author{Keith Burnett}
\affiliation{Clarendon Laboratory, University of Oxford,
  Parks Road, Oxford OX1 3PU, United Kingdom}

\date{\today}

\begin{abstract}    
We investigate the zero temperature phase diagram of a gas of bosonic
atoms in one- and two-color standing-wave lattices in the framework of the
Bose-Hubbard model.  We first introduce some relevant physical quantities;
superfluid fraction, condensate fraction, quasimomentum distribution, and
matter-wave interference pattern. We then discuss the relationships
between them on the formal level and show that the superfluid fraction,
which is the relevant order parameter for the superfluid to Mott-insulator
transition, cannot be probed directly via the matter wave interference
patterns. The formal considerations are supported by exact numerical
solutions of the Bose-Hubbard model for uniform one-dimensional systems.
We then map out the phase diagram of bosons in non-uniform lattices. The
emphasis is on optical two-color superlattices which exhibit a sinusoidal
modulation of the well depth and can be easily realized experimentally.
From the study of the superfluid fraction, the energy gap, and other
quantities we identify new zero-temperature phases, including a localized
and a quasi Bose-glass phase, and discuss prospects for their experimental
observation.   
\end{abstract}

% 03.75.Kk Dynamic properties of condensates; collective and hydrodynamic 
%	   excitations, superfluid flow
% 03.75.Lm Tunneling, Josephson effect, Bose-Einstein condensates in 
%          periodic potentials, solitons, vortices and topological excitations
% 05.30.Jp Boson systems (for static and dynamic properties of Bose-Einstein 
%          condensates, see 03.75.Hh and 03.75.Kk)
% 73.43.Nq Quantum phase transitions
% 67.40.-w Boson degeneracy and superfluidity of 4He

\maketitle

%%%%%%%%%%%%%%%%%%%%%%%%%%%%%%%%%%%%%%%%%%%%%%%%%%%%%%%%%%%%%%%%%%
%%%%%% introduction %%%%%%%%%%%%%%%%%%%%%%%%%%%%%%%%%%%%%%%%%%%%%%
%%%%%%%%%%%%%%%%%%%%%%%%%%%%%%%%%%%%%%%%%%%%%%%%%%%%%%%%%%%%%%%%%%
\section{Introduction}

With the first experimental observation of the superfluid to
Mott-insulator transition in ultracold atomic Bose gases trapped in
optical lattice potentials \cite{GrMa02a,GrMa02b} the unique prospects of
this new class of systems became evident. Compared to experiments with
Bose-Einstein condensates in shallow magnetic traps, the use of optical
standing-wave lattices gives access to a new regime which is dominated by
strong correlations. The most striking ramification of these correlations
are  quantum phase transitions \cite{Sach99} which are driven by the
interplay of the different contributions to the Hamiltonian of the system.

The superfluid to Mott-insulator transition is governed by competition
between kinetic energy and repulsive atom-atom interaction: in the
presence of strong repulsive interactions the ground state is almost a
pure Fock state with a definite occupation number at the individual
lattice sites. This is associated with a vanishing of the superfluid
density which is the natural order parameter for superfluid to insulator
phase transitions \cite{JaBr98, RoBu03a}.

New quantum phase transitions can be expected if other terms in the
Hamiltonian come into play. One interesting case is the presence of
some  additional spatial structure of the lattice potential. This might
range from a regular modulation of the well depths to random disorder.
From several studies in the context of solid-state systems it is known
that random disorder gives rise to new zero-temperature phases such as the
Anderson localized phase and the Bose-glass phase \cite{ScBa91,KrTr91}. 

Cold atoms in optical lattices offer unparalleled possibilities to study
these disorder-related phenomena \cite{RoBu03b}. By a superposition of
multiple laser beams it is possible to generate a huge variety of
different lattice topologies in a perfectly controlled manner. At the same
time Feshbach resonances can be utilized to modify the strengths of the
interatomic interaction. These powerful methods of tuning the relevant
parameters are supplemented by versatile techniques to probe the state of
the many-boson system, e.g. by observing the matter-wave interference
pattern after the atoms were released from the lattice
\cite{GrMa02a,OrTu01} or by measuring the energy gap between the ground
state and the first excited state \cite{GrMa02b}. Atoms in an optical
lattice are therefore an ideal model system for the experimental
investigation of the fundamental questions associated with quantum phase
transitions in lattice systems.

The aim of this paper is twofold: Firstly, we want to review the important
physical quantities which describe the properties of Bose gases in optical
lattices and investigate the relations between them, both, analytically and
numerically. The emphasis in this part of the paper is on the superfluid
fraction and its connection to directly observable quantities. Secondly,
we want to systematically explore the zero-temperature phase diagram for
bosons in two-color superlattices. 

The theoretical framework for the description of strongly correlated Bose
gases in lattices is the Bose-Hubbard model \cite{JaBr98}, which is
summarized in Sec. \ref{sec:bh}. We define the fundamental quantities,
condensate fraction, quasimomentum distribution, and superfluid fraction,
and discuss their relation on the formal level in Sec. \ref{sec:sfcond}.
The connection of these fundamental properties to the experimental
observables, e.g., the matter-wave interference pattern, is investigated
in Sec. \ref{sec:exp}. The formal definition shows that the superfluid
fraction depends crucially on the excitation spectrum \cite{RoBu03a}. The
basic experimental observable, i.e. the interference pattern, only probes
the ground state of the system and cannot therefore provide full
information on the superfluid properties of the system. Using the example
of the superfluid to Mott-insulator transition we illustrate these
differences in the exact numerical solution of the Bose-Hubbard model for
a one-dimensional uniform lattice. Finally, in a Sec. \ref{sec:2col} we
extend our studies to non-uniform lattice potentials. The emphasis is on
so-called two-color superlattices which exhibit a sinusoidal variation of
the well depths \cite{RoBu03b}. We map out the zero-temperature phase
diagram as function of the interaction strength and the amplitude of the
modulation and compare the behavior of number fluctuations, condensate
fraction, superfluid fraction and energy gap. This enables us to identify
additional zero-temperature phases, i.e. a localized phase and a quasi
Bose-glass phase, and to discuss possibilities for their experimental
detection.

%%%%%%%%%%%%%%%%%%%%%%%%%%%%%%%%%%%%%%%%%%%%%%%%%%%%%%%%%%%%%%%%%%
%%%%%% bose-hubbard model %%%%%%%%%%%%%%%%%%%%%%%%%%%%%%%%%%%%%%%%
%%%%%%%%%%%%%%%%%%%%%%%%%%%%%%%%%%%%%%%%%%%%%%%%%%%%%%%%%%%%%%%%%%
\section{Bose-Hubbard model}
\label{sec:bh}

Quantum phase transitions of Bose gases in optical lattices  are
associated with complex correlations in the many-body state which go far
beyond a simple Gross-Pitaevskii-like description. A theoretical model
which is capable of describing these dominating correlations is the
Bose-Hubbard model. First introduced for model studies related to
${}^{4}$He liquids in porous media and granular superconductors
\cite{FiWe89,BaSc90} it was recently applied to ultracold atomic Bose 
gases in optical lattices \cite{JaBr98}.  

In the following we will briefly review the assumptions which lead to the
Bose-Hubbard Hamiltonian and summarize solution methods and the basic
observables. Because our emphasis is on conceptual and general aspects we
restrict ourselves to one-dimensional lattices, the formal generalization
to two or three dimensions is straightforward.

%%%%%%%%%%%%%%%%%%%%%%%%%%%%%%%%%%%%%%%%%%%%%%%%%%%%%%%%%%%%%%%%%%
\subsection{Bose-Hubbard Hamiltonian}

The basic assumption of the Bose-Hubbard model is that the lattice wells
are sufficiently deep for the state of the system to be described using a
basis of single-particle wavefunctions, localized at the individual
lattice sites. Only these localized ground states are taken into account
and excited vibrational states are neglected. In the language of band
structure theory the model space of the Bose-Hubbard model comprises the
lowest energy band only, all excited bands are excluded. The localized
single-particles states are given by the Wannier functions for the lowest
band. The energy gap between the ground state band and the first excited
band has therefore to be sufficiently large for admixtures from the
excited bands to be negligible. 

A natural way to characterize the many-boson states in the model space of
the Bose-Hubbard model is the occupation number representation. Let's
assume we are dealing with a system of $N$ bosons in a lattice composed of
$I$ lattice sites. We introduce a set of occupation numbers
$\{n_1,...,n_I\}$ which specify the number of bosons in the localized
single-particle state at the individual lattice sites. The set of Fock
states $\ket{n_1,...,n_I}$ for all possible compositions of the occupation
numbers under the constraint $\sum_{i=1}^{I} n_i = N$ forms a complete
basis the Bose-Hubbard model space.

We can define the associated creation and annihilation operators, $\aaO_i$
and $\aO_i$, for a boson localized at the $i$th lattice site:
\eqmulti{
  \aaO_i \ket{n_1,...,n_i,...,n_I} 
  &= \sqrt{n_i+1}\, \ket{n_1,...,n_i+1,...,n_I} \\  
  \aO_i \ket{n_1,...,n_i,...,n_I} 
  &= \sqrt{n_i}\, \ket{n_1,...,n_i-1,...,n_I} \;. 
}
The occupation number operator for the $i$th site is given by
\eq{
  \nO_i = \aaO_i \aO_i \;.
}
Throughout this paper the indices $i,j = 1,...,I$ label the
lattice sites. All operator-valued quantities are marked by a hat. 

Using these creation and annihilation operators one can easily translate
the many-body Hamiltonian of the system---consisting of the kinetic
energy, the external lattice potential, and the two-body
interaction---into its second quantized form. This procedure leads
directly to the Bose-Hubbard Hamiltonian for the one-dimensional lattice
system  
\eqmulti{ \label{eq:hamiltonian}
  \HO_0 
  =& -J \sum_{i=1}^{I} (\aaO_{i+1} \aO_{i} + \text{h.a.}) \\
  &+ \sum_{i=1}^{I} \epsilon_i\; \nO_i
     + \frac{V}{2} \sum_{i=1}^{I} \nO_i (\nO_i-1) \;. 
}
The first term describes the tunneling between adjacent lattice sites
characterized by a strength parameter $J$. It is associated with the
kinetic energy part of the first quantized Hamiltonian. In general we will
use periodic boundary conditions, i.e., tunneling between the first and
the last lattice site is included (the site index $I+1$ which appears in
the summation is implicitly replaced by $1$). 

The second term in \eqref{eq:hamiltonian} gives an on-site single-particle
energy which originates from the external potentials and the on-site part
of the kinetic energy. The on-site energies $\epsilon_i$ are constant for
a translationally invariant lattice and might be set to zero. However, in
the presence of an additional parabolic trapping potential or for
irregular lattices the $\epsilon_i$ vary with the site index $i$. Finally,
the third term comprises the on-site two-body interaction characterized by
the interaction strength $V$. 

The parameters $J$, $\epsilon_i$, and $V$ are given by the matrix elements
of the terms of the first-quantized Hamiltonian the in the localized
single particle states \cite{JaBr98,OoSt01}. In the following we will use
the tunneling parameter $J$ as the unit of energy; the dimensionless
ratios $V/J$ and $\epsilon_i/J$ are varied to explore different regions of
the zero-temperature phase diagram of the lattice system. As mentioned in
the introduction one can envisage being able to tune the parameters of the
Hamiltonian \eqref{eq:hamiltonian} experimentally, either by changing the
lattice geometry or by exploiting a Feshbach resonance.

Besides the fundamental restriction of the Bose-Hubbard model to the
ground state band some additional assumptions are used to construct the
Hamiltonian \eqref{eq:hamiltonian}: (\emph{i}) Only tunneling between
adjacent sites is included, long range hopping over several lattice sites
is neglected. (\emph{ii}) Only short-range on-site interactions are
included; long-range interactions as they can occur for example in the
presence of dipole-dipole forces are not considered here. (\emph{iii}) The
tunneling strength $J$ is assumed to be independent of the site index.
This is obviously the case in regular lattice potentials, but it is an
approximation for irregular lattices \cite{DaZa03}. If required all these
simplifications can be easily abandoned.

%%%%%%%%%%%%%%%%%%%%%%%%%%%%%%%%%%%%%%%%%%%%%%%%%%%%%%%%%%%%%%%%%%
\subsection{Exact solution}

The most direct way to obtain the ground state and the excitation spectrum
of a zero-temperature Bose gas in one-dimensional lattices is through the
exact numerical diagonalization of the Bose-Hubbard Hamiltonian. Within
the occupation number basis the construction of the matrix representation
of the Bose-Hubbard Hamiltonian is trivial. The only substantial hurdle is
the dimension of the number basis. For fixed number of sites $I$ and
number of particles $N$ the dimension of the number basis is given by
\eq{
  D 
  = \frac{(N+I-1)!}{N! (I-1)!} \;.
}
It grows dramatically with increasing system size: For fixed filling
factor  $N/I=1$ the basis dimension for $I=8$ is $D=6435$, for $I=10$ it
grows to $D=92378$, and for $I=12$ it reaches $D=1352078$. The Hamilton
matrix is, however, extremely sparse because only the tunneling term in
\eqref{eq:hamiltonian} generates off-diagonal matrix elements. Efficient 
iterative Lanczos-type algorithms can therefore be utilized to determine
the few lowest eigenvalues and the corresponding eigenvectors. This allows
to treat systems with up to $I=N=12$ on a standard PC. 

As the result of the numerical solution of the eigenvalue problem we
obtain the energy eigenvalues $E^{(\nu)}$ and the corresponding
eigenvectors $C^{(\nu)}_{\alpha}$ for a few states ($\nu=0,1,2...$). The
eigenvectors provide the coefficients of the expansion of the eigenstates
in the number basis
\eq{ \label{eq:state}
  \ket{\Psi_{\nu}}
  = \sum_{\alpha=1}^{D} C^{(\nu)}_{\alpha}
    \ket{ \{n_1,...,n_I\}_{\alpha} } \;.
}
Here the index $\alpha=1,...,D$ labels the different Fock states, i.e.,
the different sets $\{n_1,...,n_I\}_{\alpha}$ of occupation numbers. 

Besides the exact numerical diagonalization, which explicitly yields the
interesting state vectors, it is possible to compute selected observables
using Monte-Carlo simulations \cite{BaSc92} which allow to treat systems
with several hundred lattices sites \cite{KaPr02, BaRo02}.  Because of its
simplicity and transparency we will, nevertheless, restrict ourselves to
the exact diagonalization for relatively small lattices, which already
exhibit the important universal properties of this class of systems.

Several approximation methods have been developed to solve the
Bose-Hubbard model. One of them is the so-called mean-field approximation
where the state is assumed to be a direct product of independent
single-site states. Thereby complex correlations between the individual
sites that can be of particular  importance for the description of quantum
phase transition cannot be described. On top of this independent-site
approximation the Gutzwiller ansatz \cite{KrCa92} can be used to perform
variational calculations of the ground state. 

Another approximation scheme uses the Bogoliubov approach. Here the
annihilation operators $\aO_i$ are replaced by a complex amplitude $z_i$
plus a fluctuation operator $\deltaO_i$. The fluctuation part is assumed
to be small such that terms involving squares of the fluctuation operators
can be neglected. The application and the limitations of this approach are
discussed in \cite{ReBu03}.

%%%%%%%%%%%%%%%%%%%%%%%%%%%%%%%%%%%%%%%%%%%%%%%%%%%%%%%%%%%%%%%%%%
\subsection{Simple observables}
\label{sec:simpleobs}

From the ground state $\ket{\Psi_0}$ and a few excited states obtained
from the exact diagonalization of the Bose-Hubbard Hamiltonian one can
extract several simple observables by computing expectation values.

The simplest observable is the mean occupation number of the different
sites
\eq{
  \bar{n}_i = \matrixe{\Psi_0}{\nO_i}{\Psi_0} \;.
}
For a translationally invariant lattice $\bar{n}_i$ will be the same for
all lattice sites and equal to the filling factor $N/I$ independent of the
parameters $V$ and $J$. This, however, does not mean that the structure of
the state does not change. A more telling quantity are the fluctuations
around this mean occupation number given by 
\eq{
  \sigma_i 
  = \sqrt{ \matrixe{\Psi_0}{\nO_i^2}{\Psi_0} -
    \matrixe{\Psi_0}{\nO_i}{\Psi_0}^2 } \;.
}
These number fluctuations provide direct information on the structure of
the ground state \eqref{eq:state}. If $\ket{\Psi_0}$ is a superposition of
many Fock states $\ket{ \{n_1,...,n_I\}_{\alpha} }$ then many
different occupation numbers occur at the individual sites and the
number fluctuations $\sigma_i$ will be large. If, on the other hand, the
ground state is described by a single Fock state, then the fluctuations
will vanish. 

As a complementary measure the magnitude of the largest coefficient in the
Fock state expansion \eqref{eq:state} of the ground state can be used
\eq{
  C^2_{\max} 
  = \max(C_{\alpha}^2) \;.
}  
Although this is not directly observable it yields sensitive information on
the structure of the state. A small value of $C^2_{\max}$ indicates that
$\ket{\Psi_0}$ is a superposition of many number states;
$C^2_{\max}\approx1$ means that the ground state is a pure Fock state.
 
%%%%% figure %%%%%%%%%%%%%%%%%%%%%%%%%%%%%%%%%%%%%%%%%%%%%%%%%%%%%
\begin{figure}
\includegraphics[width=0.37\textwidth]{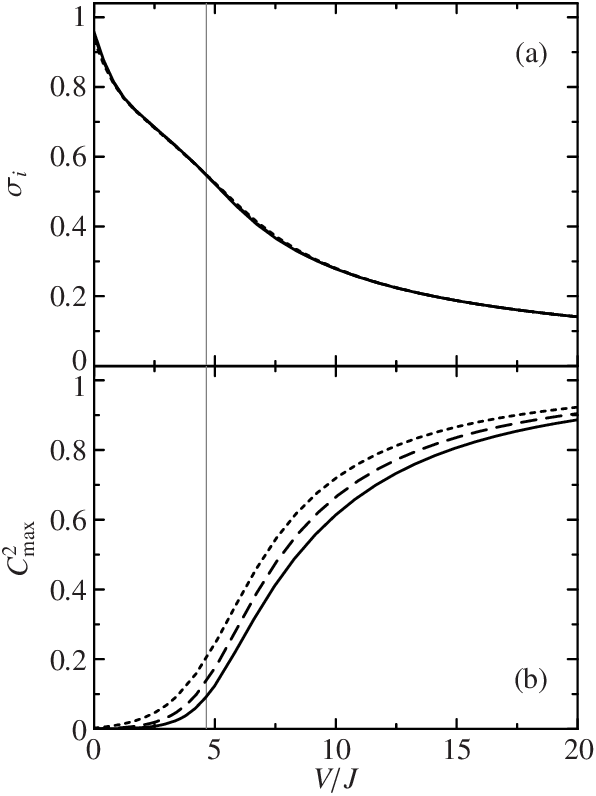}
\caption{Number fluctuations $\sigma_i$ (a) and largest coefficient
$C_{\max}^2$ (b) for the ground state of the one-dimensional Bose-Hubbard
model with $I=N=12$ (solid), $10$ (dashed), and $8$ (dotted) as function
of the interaction strength $V/J$. The vertical gray line marks the
critical interaction strength for the Mott-insulator transition in an
infinite system \cite{BaSc92,FrMo96}.}
\label{fig:mott_nfluc_cmax}
\end{figure}
%%%%%%%%%%%%%%%%%%%%%%%%%%%%%%%%%%%%%%%%%%%%%%%%%%%%%%%%%%%%%%%%%%

To illustrate these quantities we solve the eigenvalue problem of the
Bose-Hubbard Hamiltonian \eqref{eq:hamiltonian} for a regular lattice
potential ($\epsilon_i=0$) with filling factor $N/I=1$ for different
ratios $V/J$ of the interaction strength and the tunneling coefficient.
The results for the number fluctuations $\sigma_i$ and the largest
coefficient $C_{\max}^2$ are depicted in Fig. \ref{fig:mott_nfluc_cmax}.
The vertical gray line indicates the critical interaction strength
$(V/J)_{\text{cr}}=4.65$ for the superfluid to Mott-insulator transition
in an infinite one-dimensional lattice extracted from a Monte Carlo
calculation \cite{BaSc92} and from a strong coupling expansion
\cite{FrMo96}. It marks the region where structural changes in the ground
state are also expected in the finite size systems considered here.

Our first important observation is that the number fluctuations decrease
rather slowly with increasing interaction strength. Moreover, there
appears no clear signature of a phase transition in their variation. This
is not an artifact of the limited lattice size, because the number
fluctuation exhibit no noticeable size dependence as the three curves for
$I=12,10,$ and $8$ in Fig. \ref{fig:mott_nfluc_cmax}(a) illustrate. Thus
we conclude that the number fluctuations and related quantities do not
reveal the quantum phase transition from a superfluid to a Mott-insulator
state and that the insulating state still has significant number
fluctuations, i.e., is not a pure Fock state.

The last point is confirmed by the behavior of the largest coefficient
$C_{\max}^2$ plotted in Fig. \ref{fig:mott_nfluc_cmax}(b). This quantity
is very small in the insulating phase and starts to grow in the region of
the phase transition. However, even for large values of the ratio $V/J$
the largest coefficient in \eqref{eq:state} remains smaller than $1$
indicating that there is a dominant Fock state (the one with $n_i=N/I$)
but that other Fock states also contribute to the ground state.

%%%%% figure %%%%%%%%%%%%%%%%%%%%%%%%%%%%%%%%%%%%%%%%%%%%%%%%%%%%%
\begin{figure}
\includegraphics[width=0.37\textwidth]{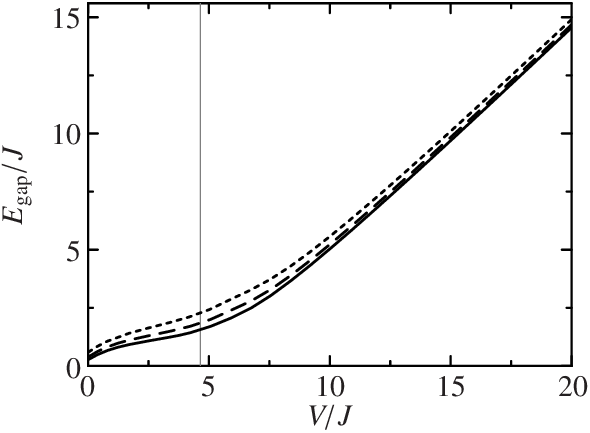}
\caption{Energy gap $E_{\gap}$ between ground state and first excited state
as function of the interaction strength $V/J$ for commensurate filling
$N/I=1$ and different lattice sizes $I=12$ (solid line), $10$ (dashed),
$8$ (dotted).}
\label{fig:mott_egap}
\end{figure}
%%%%%%%%%%%%%%%%%%%%%%%%%%%%%%%%%%%%%%%%%%%%%%%%%%%%%%%%%%%%%%%%%%

Another important observable is the energy gap, i.e., the difference
between the energies of the ground state and the first excited state, 
\eq{
  E_{\gap} = E^{(1)} - E^{(0)} \;.
}
This quantity is of particular importance as the excited states are
crucial for the superfluid properties of the system. Figure
\ref{fig:mott_egap} depicts the behavior of the energy gap as function of
the interaction strength. For small $V/J$ the energy gap is small and
increases only slowly with increasing interaction strengths. The slope
then starts to increase until, for $V/J\gtrsim10$, one observes a linear
increase of the energy gap with a slope $1$, i.e., changing the
interaction strength $V$ by a certain amount will change the energy gap by
the same amount. Qualitatively, this behavior can be explained by assuming
that the excited state is a superposition of different one-particle
one-hole excitations of the ground state. The predominant Fock states in
the expansion \eqref{eq:state} of the excited state thus have one doubly
occupied site which leads to the linear dependence of the energy gap
$E_{\gap}$ on $V$. These signatures of the Mott-insulator phase have been
investigated experimentally \cite{GrMa02b} as well as theoretically 
\cite{SaSe02,BrDu02} by tilting the lattice potential. 

The behavior of the energy gap indicates that the character of the energy
spectrum changes in the region $V/J\sim 5$ to $8$. As we will see in Sec.
\ref{sec:sf} this change is accompanied by a vanishing of superfluidity,
i.e. the superfluid to Mott-insulator transition.

%%%%%%%%%%%%%%%%%%%%%%%%%%%%%%%%%%%%%%%%%%%%%%%%%%%%%%%%%%%%%%%%%%
%%%%%% condensate and superfluidity %%%%%%%%%%%%%%%%%%%%%%%%%%%%%%
%%%%%%%%%%%%%%%%%%%%%%%%%%%%%%%%%%%%%%%%%%%%%%%%%%%%%%%%%%%%%%%%%%
\section{Condensate and superfluidity}
\label{sec:sfcond}

Besides the simple quantities discussed so far there are two more complex
physical quantities which reveal substantial information on fundamental
physical properties of the Bose gas in the lattice. These are the
condensate fraction and the superfluid fraction. This section is devoted to
the definition and discussion of these non-trivial quantities.

%%%%%%%%%%%%%%%%%%%%%%%%%%%%%%%%%%%%%%%%%%%%%%%%%%%%%%%%%%%%%%%%%%
\subsection{Condensate fraction}

Unlike the Gross-Pitaevskii description of weakly interacting Bose gases 
\cite{DaGi99} the exact solution of the Bose-Hubbard model does not assume
a perfect Bose-Einstein condensate from the outset. Moreover, due to the
general  representation of the ground state in terms of Fock states the
presence or absence of a Bose-Einstein condensate is not obvious. For
example, the exact ground state \eqref{eq:state} of a noninteracting
system at zero temperature, where one expects a perfect Bose-Einstein
condensate, is a superposition of many Fock states with coefficients given
by the  multinomial distribution. 

%%%%%%%%%%%%%%%%%%%%%%%%%%%%%%%%%%%%%%%%%%%%%%%%%%%%%%%%%%%%%%%%%%
\subsubsection{Definition}

In order to define a Bose-Einstein condensate for the ground state
$\ket{\Psi_0}$ of a general interacting many-boson system in a lattice we
consider the one-body density matrix associated with $\ket{\Psi_0}$
\eq{ \label{eq:cond_densitymat}
  \rho^{(1)}_{ij} 
  = \matrixe{\Psi_0}{\aaO_j\aO_i}{\Psi_0} \;.
}
The eigenvectors of the one-body density matrix describe the so-called
natural orbitals and the eigenvalues the corresponding occupation
numbers. 

Following the formulation of Penrose and Onsager \cite{PeOn56} a
Bose-Einstein condensate is present if one of the natural orbitals is
\emph{macroscopically} occupied. Its occupation number is just the 
number of condensate particles $N_{\cond}$ and the eigenvector
constitutes the condensate wave function $\chi_{\cond,i}$. 

Strictly speaking, macroscopic occupation of an orbital means that the
ratio of its occupation number and the total particle number, in the
following called condensate fraction $f_{\cond}=N_{\cond}/N$, remains
finite in the the thermodynamic limit ($N,I \to\infty$,
$N/I=\text{const.}$). This implies that one cannot rigorously
determine the \emph{absence} of a Bose-Einstein condensate in
finite-size systems. From the normalization of the one-body density
matrix, $\mathrm{Tr}\,\rho^{(1)}_{ij} = N$, it follows immediately that
for any system there is an eigenvalue larger or equal $N/I$. Thus the
condensate fraction $f_{\cond}$ is always larger than $1/I$ and cannot
vanish for finite-size systems. Despite of these formal complications the
ratio of the largest eigenvalue of \eqref{eq:cond_densitymat} and the
total particle number provides important physical information even in
finite lattices. For simplicity we will use the term condensate fraction
for this quantity.

The presence of a Bose-Einstein condensate implies off-diagonal long range
order \cite{Yang62}. That is, the matrix elements of the one-body density
matrix do not go to zero far off the diagonal but remain finite:
\eq{
  \rho^{(1)}_{ij} \nrightarrow 0 \quad\text{for}\quad |i-j| \to \infty \;.
}
The prove of this connection is straightforward if one considers the
spectral decomposition of the one-body density matrix which allows us to
separate the condensate part from a residual positive semi-definite density
matrix $\tilde{\rho}^{(1)}_{ij}$
\eq{
  \rho^{(1)}_{ij}
  = N_{\cond}\; \conj{\chi}_{\cond,i} \chi_{\cond,j}
  + \tilde{\rho}^{(1)}_{ij} \;.
}
It is important to realize that off-diagonal long range order is
associated directly with the phenomenon of Bose-Einstein condensation
whereas the connection with superfluidity is, at most, indirect.

%%%%% figure %%%%%%%%%%%%%%%%%%%%%%%%%%%%%%%%%%%%%%%%%%%%%%%%%%%%%
\begin{figure}
\includegraphics[width=0.37\textwidth]{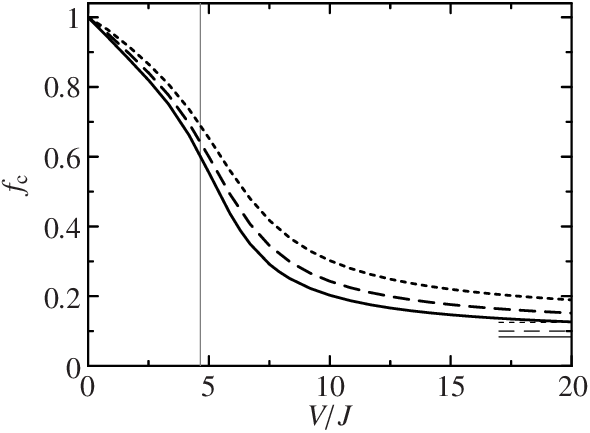}
\caption{Condensate fraction $f_{\cond}$ as function of the interaction
strength $V/J$ for commensurate filling $N/I=1$ and different lattice
sizes $I=12$ (solid line), $10$ (dashed), $8$ (dotted). The thin lines at
the right border indicate the asymptotic value $f_{\cond}\to1/I$.}
\label{fig:mott_condensate}
\end{figure}
%%%%%%%%%%%%%%%%%%%%%%%%%%%%%%%%%%%%%%%%%%%%%%%%%%%%%%%%%%%%%%%%%%

The dependence of the condensate fraction, i.e., the relative occupation of
the natural orbital with largest eigenvalue, on the interaction strength
$V/J$ in a regular lattice is depicted in Fig. \ref{fig:mott_condensate}.
In the noninteracting case $V/J=0$ all particles are in the condensate. 
As soon as a finite repulsive interaction is present ($V/J>0$) the
condensate fraction is reduced, i.e., the condensate is depleted and
non-condensate states are populated. For large interaction strengths the
condensate fraction converges to $1/I$. This is the reason for the
pronounced size dependence of the condensate fraction in the regime of
strong depletion.

%%%%%%%%%%%%%%%%%%%%%%%%%%%%%%%%%%%%%%%%%%%%%%%%%%%%%%%%%%%%%%%%%%
\subsubsection{Quasimomentum Distribution}
\label{sec:quasimomdist}

The eigenvalues of the one-body density matrix \eqref{eq:cond_densitymat}
provide more information than just the condensate fraction. For a
translationally invariant lattice one can show that the natural orbitals
satisfy the Bloch theorem and therefore are quasimomentum eigenstates
\cite{JoMa85a}. Thus the occupation numbers of the natural orbitals
correspond to occupation numbers $\tilde{n}_q$ of the Bloch states with
different quasimomenta $q$. Notice that for a finite lattice of length
$L$ the quasimomenta can assume only discrete values which are integer
multiples of $2\pi/L$. The Bloch function for quasimomentum $q=0$
describes the condensate state as defined above.

An alternative way to obtain the quasimomentum distribution for a regular
lattice is to use the relation between the localized Wannier functions
and delocalized Bloch functions, which are the quasimomentum eigenstates.
The Bloch functions $\psi_{q}(x)$ of the lowest band can be decomposed in
terms of Wannier functions $w(x-\xi_i)$: 
\eq{
  \psi_q(x) 
  = \frac{1}{\sqrt{I}} \sum_{i=1}^{I} \ee^{-\ii q \xi_i} w(x-\xi_i) \;,
}
where $\xi_i$ is the coordinate of the $i$th lattice site. Using the fact
that the creation operators $\aaO_i$ of the Bose-Hubbard model create a
boson in the Wannier state $w(x-\xi_i)$ we can readily define creation
operators $\ccO_q$ for a boson in the Bloch state with quasimomentum $q$
\cite{OoSt01}
\eq{
  \ccO_q
  = \frac{1}{\sqrt{I}} \sum_{i=1}^{I} \ee^{-\ii q \xi_i}\; \aaO_i \;.
}
The occupation numbers of the Bloch states with quasimomentum $q$ are thus
given by
\eqmulti{ \label{eq:cond_quasimomdist}
  \tilde{n}_q 
  = \matrixe{\Psi_0}{\ccO_q \cO_q}{\Psi_0} 
  &= \frac{1}{I} \sum_{i,j=1}^{I}
    \ee^{\ii\,q (\xi_i-\xi_j)}\; \matrixe{\Psi_0}{\aaO_j\aO_i}{\Psi_0}\\
  &= \frac{1}{I} \sum_{i,j=1}^{I}
    \ee^{\ii\,qa\,(i-j)}\;\rho^{(1)}_{ij} \;,
}
where $a = \xi_{i+1}-\xi_i$ is the lattice spacing. As expected, the
quasimomentum distribution is related to the Fourier transform of the
one-body density matrix.
 
%%%%% figure %%%%%%%%%%%%%%%%%%%%%%%%%%%%%%%%%%%%%%%%%%%%%%%%%%%%%
\begin{figure}
\includegraphics[width=0.37\textwidth]{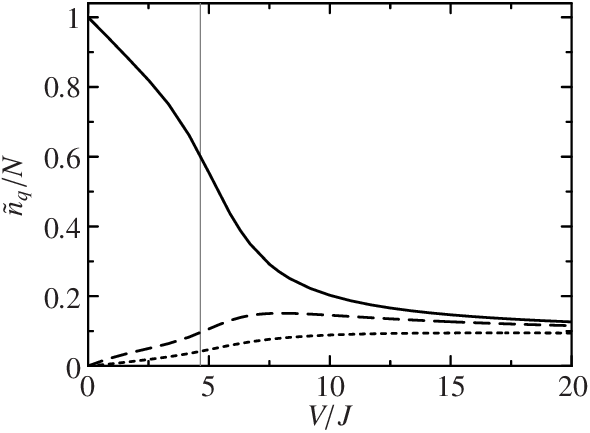}
\caption{Occupation numbers $\tilde{n}_q$ of Bloch states with
quasimomentum $q$ function of the interaction  strength $V/J$ for a
lattice with $I=N=12$. The solid line shows the condensate $q=0$ state,
the dashed and dotted lines correspond to the first and second excited
quasimomentum state, respectively.}
\label{fig:mott_quasimomdist}
\end{figure}
%%%%%%%%%%%%%%%%%%%%%%%%%%%%%%%%%%%%%%%%%%%%%%%%%%%%%%%%%%%%%%%%%%

Figure \ref{fig:mott_quasimomdist} shows the occupation number for the
three lowest quasimomentum states as function of the interaction strength
$V/J$. The population for the $q=0$ state corresponds to the condensate
fraction and is depleted as the interaction strength is increased.
Simultaneously with this, the occupation numbers of Bloch states with
higher quasimomenta, which vanish in the noninteracting system, increase
successively. The repulsive interactions thus lead to a redistribution of
the particles from the condensate to states with higher quasimomenta. In
the limit of strong interactions all quasimomentum states exhibit the
same occupation, i.e., the band is filled uniformly.

%%%%%%%%%%%%%%%%%%%%%%%%%%%%%%%%%%%%%%%%%%%%%%%%%%%%%%%%%%%%%%%%%%
\subsection{Superfluid fraction}
\label{sec:sf}

One of the most interesting physical quantities in an interacting
many-boson system is the superfluid density. It is the natural order
parameter for the various superfluid to insulator phase transitions, e.g.,
the superfluid to Mott-insulator transition observed experimentally
\cite{GrMa02a}.

%%%%%%%%%%%%%%%%%%%%%%%%%%%%%%%%%%%%%%%%%%%%%%%%%%%%%%%%%%%%%%%%%%
\subsubsection{Definition}

The macroscopic, as well as the microscopic definition of superfluidity,
are connected to flow properties. Macroscopically, one usually employs the
two-fluid picture to describe systems which exhibit superfluidity. The
superfluid and the normal fluid are distinguished through their behavior
in the presence of moving boundaries. For example, in a narrow  channel
with moving side walls only the normal fluid is dragged along with the
walls, whereas the superfluid stays at rest. In the rest frame of the
walls only the superfluid moves. From the difference between the energies
(in the rest frame of the walls) of a system with moving and one with
stationary walls, one can directly extract the kinetic energy of the
superfluid component and hence, via its velocity, the superfluid density
\cite{PoCe87,LiSe02}. 

A microscopic definition of superfluidity can be constructed in analogy to
this simple macroscopic picture. First of all we consider the generalized
condensate wave function $\chi(\xV)=\ee^{\ii \vartheta(\xV)} |\chi(\xV)|$
as defined in the previous section. A spatial variation of the phase
$\vartheta(\xV)$ of the condensate wave function is connected to a
velocity field \cite{Legg99}
\eq{ \label{eq:sf_velocity}
 \vV_{\SF} 
 = \frac{\hbar}{m}\, \nablaV \vartheta(\xV) \;.
}
The flow described by this velocity field is non-dissipative and
irrotational, properties which conform with the macroscopic picture of
superfluidity. This velocity is therefore identified with the  flow field
of a superfluid \cite{Legg99}. In principle Eq. \eqref{eq:sf_velocity}
provides the basic microscopic distinction of the superfluid through its
flow behavior.

As in the macroscopic case we can develop an explicit expression for  the
superfluid fraction by considering the total energy of the system. For
simplicity we restrict ourselves to a one-dimensional system. Assume that
we impose a linear phase variation  with a total phase twist $\Theta$ over
the length $L$ of the system, i.e., $\vartheta(x) = \Theta\, x/L$. One way
to accomplished this is by imposing twisted boundary conditions on the
many-body wave function
\eq{ \label{eq:sf_twistedbc}
  \Psi(x_1,...,x_k+L,...,x_N) 
  =  \ee^{\ii \Theta} \Psi(x_1,...,x_k,...,x_N)
}
for all $k$ \cite{FiBa73,ShSu90,Poil91}. This imposed phase gradient
generates a flow with a velocity  $v_{\SF} = \hbar \Theta/(m L)$. The
portion of the system which responds to the phase gradient by flowing with
this velocity is called superfluid. The presence of the flow will increase
the total energy of the system. As long as other excitations are absent,
that is, as long as the imposed phase gradient is small, the increase in
the total energy can be attributed solely to the kinetic energy of the
superflow
\eq{ \label{eq:sf_energy}
  E_{\Theta} - E_{0} 
  = \frac{1}{2} M_{\SF} v_{\SF}^2 \;.
}
Here $E_0$ is the ground state energy of the system without phase twist
and $E_{\Theta}$ is the ground state energy obtained with twisted boundary
conditions. The mass $M_{\SF}$ corresponds to the total mass of the
superfluid portion, by introducing the superfluid fraction $f_{\SF}$ is
can be rewritten as $M_{\SF} = m N f_{\SF}$. From Eq. \eqref{eq:sf_energy}
we can extract a simple expression for the superfluid fraction
\cite{FiBa73,Krau91,SiRo94}
\eq{ \label{eq:sf_frac_cont}
  f_{\SF}
  = \frac{2 m L^2}{\hbar^2 N}\; \frac{E_{\Theta} - E_{0}}{\Theta^2} 
  \qquad\text{for}\;\Theta\ll\pi \;.
}
Hence the superfluid fraction is determined by the stiffness of the system
under phase variations. We have to stress that the twist angle $\Theta$
has to be sufficiently small to avoid effects other than the collective
flow of the superfluid component. We will return to this point at the end
of this section.

The definition of the superfluid fraction through the energy change under
imposed phase variations corresponds to the so-called helicity modulus
introduced in Ref. \cite{FiBa73}. Moreover, it is equivalent to the concept
of winding numbers which is used in path-integral Monte Carlo methods
\cite{PoCe87}. All these measures predict the noninteracting Bose gas to
be a perfect superfluid. This is in contradiction to the Landau picture of
superfluidity which is based on the dispersion relation of elementary
excitations which defines a critical velocity. In the Landau picture the
ideal Bose gas is not considered a superfluid since its critical velocity
is zero. A possible connection between these two pictures of superfluidity
is still a matter of debate \cite{LiSe02}. We adopt a pragmatic view and
use the definition \eqref{eq:sf_frac_cont} keeping in mind that it does
not tell anything about the dynamical stability of the superfluid flow at
finite velocities. 

A second comment is appropriate concerning the relation between
superfluidity and Bose-Einstein condensation. First of all  one should
realize that although the phase of the condensate wave function determines
the superfluid velocity \eqref{eq:sf_velocity}, the superfluid and the
condensate fraction are not the same. A famous example is liquid ${}^4$He
at zero temperature which is 100\% superfluid but only 10\% of the atoms
are in the condensate \cite{Legg99}. From the formal point of view we used
the condensate wave function only to motivate the connection between
twisted boundary conditions \eqref{eq:sf_twistedbc} and the superfluid
flow. On the level of the resulting Eqs. \eqref{eq:sf_twistedbc} and
\eqref{eq:sf_frac_cont} the condensate does not appear. In general it is not
strictly evident that superfluidity presupposes the presence of a
condensate \cite{LiSe02}. 

The relation \eqref{eq:sf_frac_cont} for the continuous  system can be
easily transfered to the Bose-Hubbard model. Basically one replaces the
length $L$ of the continuous system by the number of lattice sites $I$ and
the prefactor of the kinetic energy $\hbar^2/(2m)$ by the tunneling
strength $J$ of the Bose-Hubbard Hamiltonian. The superfluid fraction for
the discrete lattice thus reads \cite{RoBu03a}
\eq{ \label{eq:sf_frac}
  f_{\SF}
  = \frac{I^2}{J N} \; \frac{E_{\Theta} - E_{0}}{\Theta^2}
  \qquad\text{for}\;\Theta\ll\pi\;.
}

In order to compute the ground state energy of a Bose-Hubbard system with
imposed phase gradient, the use of twisted boundary conditions is
impracticable. By means of the local unitary transformation $\UO_{\Theta} =
\prod_{k=1}^{N}\exp(\ii\Theta\, \xO_k/L)$ we can map the phase
twist from the state onto the Hamiltonian. The eigenvalues of the
resulting twisted Hamiltonian for periodic boundary conditions are
identical to those of the original Hamiltonian with twisted boundary
conditions. The twisted Bose-Hubbard Hamiltonian reads
\eqmulti{ \label{eq:hamiltonian_twist}
  \HO_{\Theta} 
  =& -J \sum_{i=1}^{I} (\ee^{-\ii \Theta/I}\, \aaO_{i+1} \aO_{i} 
    + \text{h.a.}) \\ 
  &+ \sum_{i=1}^{I} \epsilon_i\; \nO_i
    + \frac{V}{2} \sum_{i=1}^{I} \nO_i (\nO_i-1) \;. 
}
Compared to the original Hamiltonian \eqref{eq:hamiltonian} only the
off-diagonal hopping term is modified by an additional factor
$\ee^{\mp\ii \Theta/I}$, the so-called Peierls phase factor
\cite{ShSu90,Poil91}. Physically this term ensures that any particle which
tunnels to an adjacent site acquires the correct phase in order to
establish the linear phase variation across the system.

There are several ways to impose this phase factor experimentally. One
possibility is to add a linear external potential or a homogeneous
electric field in the case of charged particles. Another possibility is a
constant acceleration of the lattice, which in the case of optical
standing wave lattices can be realized by ramping the detuning between the
two counter-propagating beams. One might be able to utilize this
possibility to probe superfluidity directly. 

A straightforward procedure to compute the superfluid fraction is now
evident: We numerically solve the eigenvalue problems of the twisted and
the non-twisted Hamiltonian, $\HO_{\Theta}$ and $\HO_{0}$, with periodic
boundary conditions and insert the resulting ground state energies
$E_{\Theta}$ and $E_{0}$ into \eqref{eq:sf_frac}.

%%%%% figure %%%%%%%%%%%%%%%%%%%%%%%%%%%%%%%%%%%%%%%%%%%%%%%%%%%%%
\begin{figure}
\includegraphics[width=0.37\textwidth]{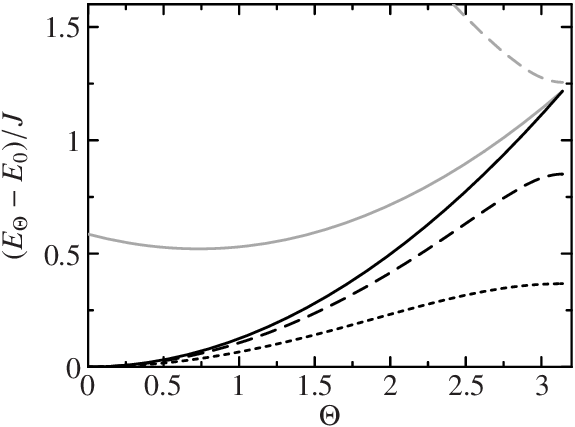}
\caption{Difference $E_{\Theta}-E_0$ between the ground state energies with
an imposed phase twist and without as function of the total twist angle
$\Theta$ for a system with $I=N=8$. The different curves correspond to
$V/J=0$ (solid line), $V/J=3$ (dashed), and $V/J=5$ (dotted). The gray
lines depict the energy difference for the first excited state of the 
twisted system.}
\label{fig:energy-vs-theta}
\end{figure}
%%%%%%%%%%%%%%%%%%%%%%%%%%%%%%%%%%%%%%%%%%%%%%%%%%%%%%%%%%%%%%%%%%

Before doing so, it is useful to study how the energy difference
$E_{\Theta}-E_{0}$ entering into Eq. \eqref{eq:sf_frac} depends on the
twist angle $\Theta$.  We have already assumed that $\Theta$ has to be
sufficiently small to avoid excitations other than the collective flow of
the superfluid component. Figure \ref{fig:energy-vs-theta} depicts the
result of a numerical calculation for $E_{\Theta}-E_{0}$ as function of
$\Theta$ for different values of $V/J$. For the noninteracting system
(black solid curve) the energy is proportional to $\Theta^2$ up to large
values of $\Theta$, i.e., the superfluid fraction is independent of the
actual value of $\Theta$ used in the calculation.  However, one should
note that at $\Theta=\pi$, which corresponds to antiperiodic boundary
conditions, a level crossing between the ground state and the first
excited state of the twisted system appears. The presence of interactions
leads to a ``level repulsion'' between these two states (see dashed and
dotted curves in Fig. \ref{fig:energy-vs-theta}). Hence the energy
deviates significantly from the $E_{\Theta}-E_0 \propto \Theta^2$ behavior
for $\Theta\sim\pi$. If one were to use Eq. \eqref{eq:sf_frac} for
$\Theta\sim\pi$ one would underestimate the superfluid fraction
significantly. A physically meaningful application of Eq.
\eqref{eq:sf_frac} to determine the superfluid fraction therefore requires
$\Theta\ll\pi$. Formally one might even impose the limit $\Theta\to0$. For
all the following numerical calculations we will use $\Theta=0.1$.

%%%%%%%%%%%%%%%%%%%%%%%%%%%%%%%%%%%%%%%%%%%%%%%%%%%%%%%%%%%%%%%%%%
\subsubsection{Direct numerical calculation}
 
%%%%% figure %%%%%%%%%%%%%%%%%%%%%%%%%%%%%%%%%%%%%%%%%%%%%%%%%%%%%
\begin{figure}
\includegraphics[width=0.37\textwidth]{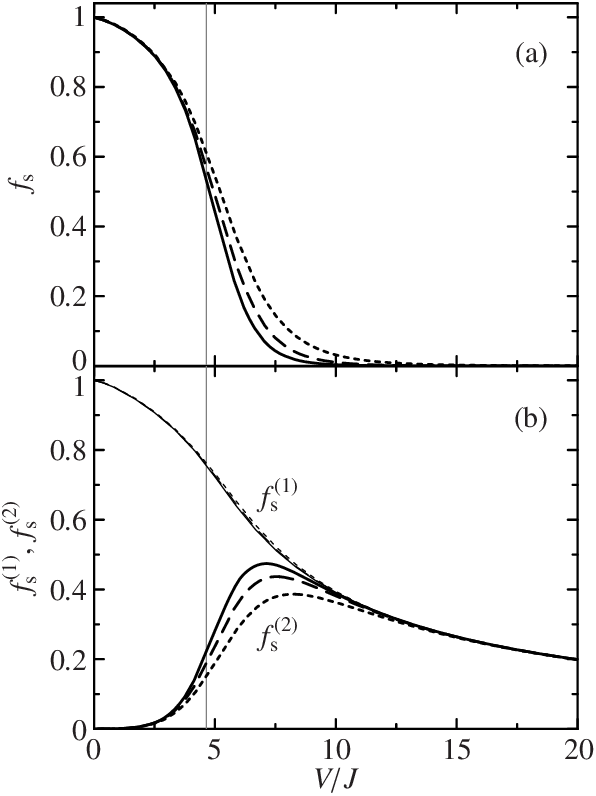}
\caption{(a) Superfluid fraction $f_{\SF}$ for one-dimensional Bose-Hubbard
systems with $I=N=12$ (solid), $10$ (dashed), and $8$ (dotted) as function
of the interaction strength $V/J$. (b) First and second order contribution
to $f_{\SF}$ in the perturbative calculation. The vertical gray line marks
the critical interaction strength $(V/J)_{\text{cr}}=4.65$ for the
Mott-insulator transition in an infinite system \cite{BaSc92,FrMo96}.}
\label{fig:mott_superfluid}
\end{figure}
%%%%%%%%%%%%%%%%%%%%%%%%%%%%%%%%%%%%%%%%%%%%%%%%%%%%%%%%%%%%%%%%%%

We now employ Eq. \eqref{eq:sf_frac} to compute the superfluid fraction
$f_{\SF}$ from the energies $E_{\Theta}$ and $E_{0}$ obtained numerically.
The results for systems with commensurate filling $N/I=1$ and lattice
sizes up to $I=12$ are shown in Fig. \ref{fig:mott_superfluid}(a). 

In the range $3 < V/J < 6$ the superfluid fraction, i.e. the order
parameter for the superfluid to Mott-insulator transition, declines
rapidly with increasing interaction strength $V/J$. In these relatively
small systems we already observe a rapid transition from the superfluid
phase at small $V/J$ to an insulating phase with vanishing superfluid
fraction  at large $V/J$. The transition region is in good agreement with
the critical point $(V/J)_{\text{cr}}=4.65$ for the infinite lattice
determined by Monte Carlo simulations \cite{BaSc92} or a strong coupling
expansion \cite{FrMo96}.

The dependence of the superfluid fraction on the lattice size shown in
Fig. \ref{fig:mott_superfluid}(a) is rather moderate. In the superfluid
phase for up to $V/J\leq4$ there is no noticeable size dependence. Only the
shoulder of the curve exhibits a size dependence: With increasing $I$ the
decrease in $f_{\SF}$ steepens and the tail of the curve is compressed and
shifted towards lower $V/J$. From this observation we conclude that
systems with $I\approx10$ already allow very good qualitative, and even
semi-quantitative conclusions, about the superfluid behavior of larger
systems.

The rapid decrease of the superfluid fraction is in stark contrast to the
behavior of the number fluctuations shown in Fig.
\ref{fig:mott_nfluc_cmax}(a). Evidently, the number fluctuations still
exhibit values of up to $\sigma_i\approx 0.3$ in cases where the
superfluid fraction has already vanished. This proves that the
Mott-insulator state can still be a superposition of many Fock states and
thus show considerable number fluctuations. Only in the limit
$V/J\to\infty$ when $\sigma_i$ vanishes it becomes a pure Fock state.

%%%%%%%%%%%%%%%%%%%%%%%%%%%%%%%%%%%%%%%%%%%%%%%%%%%%%%%%%%%%%%%%%%
\subsubsection{Perturbative treatment}
\label{sec:superfluid_3}

Apart from the direct solution of the eigenvalue problems of the twisted
and the non-twisted Hamiltonian one can use a perturbative approach to
determine the energy difference $E_{\Theta}-E_0$. As the superfluid
fraction \eqref{eq:sf_frac} is defined only for very small twist angles
$\Theta$ this perturbative treatment does not introduce any additional
approximation. All terms of the expansion that contribute in the limit
$\Theta\to0$ will be included.

The derivation consists of two steps: We first expand the twisted
Hamiltonian in a power series up to second order in $\Theta$
\eq{ \label{eq:hamiltonain_twistexp}
  \HO_{\Theta}
  \simeq \HO_0 + \frac{\Theta}{I}\, \JO - \frac{\Theta^2}{2 I^2}\, \TO
  = \HO_0 + \HO_{\text{pert}} \;,
}
where the current operator $\JO$ and the kinetic energy
operator $\TO$ are given by:
\eqmulti{
  \JO &= \ii J \sum_{i=1}^{I} (\aaO_{i+1} \aO_{i} - \text{h.a.}) \\
  \TO &= -J \sum_{i=1}^{I} (\aaO_{i+1} \aO_{i} + \text{h.a.}) \;.
}
We then compute the energy shift $E_{\Theta} - E_0$ caused by the
perturbation $\HO_{\text{pert}}$ in \eqref{eq:hamiltonain_twistexp} via
second order perturbation theory. The lowest order contributions to the
energy shift are quadratic in $\Theta$. Inserting them into
\eqref{eq:sf_frac} leads to an expression for the superfluid fraction 
consisting of two terms
\eq{ \label{eq:sf_frac_drude}
  f_{\SF} 
  = f_{\SF}^{(1)} - f_{\SF}^{(2)} \;.
}
The first order term is proportional to the kinetic energy expectation
value
\eq{ \label{eq:sf_frac_drude1}
  f_{\SF}^{(1)}
  = -\frac{1}{2NJ} \matrixe{\Psi_0}{\TO}{\Psi_0} 
  = \frac{1}{N} \sum_{i=1}^{I} \rho^{(1)}_{i,i+1}\;,
}
where $\Psi_0$ is the ground state of the original Hamiltonian $\HO_0$ and
$\rho^{(1)}_{ij}$ the associated one-body density matrix introduced in Eq.
\eqref{eq:cond_densitymat}. The second order term involves matrix elements
of the current operator between the ground state $\ket{\Psi_0}$ and all
the excited states $\ket{\Psi_{\nu}}$ ($\nu=1,2,...$) of the original
Hamiltonian $\HO_0$,
\eq{ \label{eq:sf_frac_drude2}
  f_{\SF}^{(2)}
  = \frac{1}{NJ} \sum_{\nu\ne0} 
  \frac{|\matrixe{\Psi_{\nu}}{\JO}{\Psi_0}|^2}{E^{(\nu)} - E^{(0)}} \;.
}
All higher-order terms of the perturbative expansion of the energy shift
do not contribute to the superfluid fraction in the limit $\Theta\to0$.
Thus Eq. \eqref{eq:sf_frac_drude} is an exact expression for
$f_{\SF}$.

We note that this derivation and the resulting expression is closely
related to the so-called Drude weight which characterizes the dc
conductivity of charged fermionic systems \cite{FyMa91}. 

This formulation of the superfluid fraction provides a detailed insight
into the mechanisms that govern the superfluid properties of the system.
First of all it shows that superfluidity is not a property of the ground
state, but rather the response of the system to an external perturbation.
The second order term \eqref{eq:sf_frac_drude2} introduces an explicit
dependence on the whole excitation spectrum. Since this second order
contribution always lowers the total superfluid fraction we can conclude
that the coupling to excited states leads to a suppression of
superfluidity. The first order term \eqref{eq:sf_frac_drude1}, which is
just the ground state expectation value of the kinetic energy and can be
expressed in terms of the one-body density matrix of the ground state
alone, gives only an upper bound for $f_{\SF}$ \cite{PaTr98}. 

The behavior of the first and second order contributions to the superfluid
fraction for a regular lattice are shown in Fig.
\ref{fig:mott_superfluid}(b). The first order term \eqref{eq:sf_frac_drude1},
i.e. the rescaled kinetic energy expectation value, is approximately one
for small $V/J$ and decreases slowly with increasing interaction strength.
Within the insulating phase, where the total superfluid fraction shown in
Fig. \ref{fig:mott_superfluid}(a) has vanished completely, the first order
contribution $f_{\SF}^{(1)}$ still has values of up to $0.3$.

The vanishing of the total superfluid fraction is caused by a 
characteristic behavior of the second order contribution
\eqref{eq:sf_frac_drude2}: In the transition region  $f_{\SF}^{(2)}$ shows
a threshold-like increase from zero to its maximum value. For values of
$V/J$ beyond this maximum the second order term quickly converges to the
value of $f_{\SF}^{(1)}$. Thus the vanishing of the superfluid fraction in
insulating phases, e.g., the Mott-insulator, results from the cancellation
between first and second order term which both have considerable size.

This clearly demonstrates that the coupling to the excited states and the
structure of the excitation spectrum is essential for the superfluid
properties of the system. The ground state alone, represented by the first
order contribution $f_{\SF}^{(1)}$, provides only limited
information---namely an upper bound---on the superfluid fraction.
Therefore, any observable which is sensitive to the ground state only,
such as number fluctuations, condensate fraction, and interference
pattern, cannot provide clear information the superfluid properties of the
system.

The only other observable which is sensitive to the properties of the
excitation spectrum is the energy gap $E_{\gap}$ between the ground state
and the first excited state. As we have shown in Sec. \ref{sec:simpleobs}
the energy gap starts to grow linearly with $V/J$ in the Mott-insulator
phase. This is another indicator for the change in the excitation spectrum
that also leads to the vanishing of the superfluid fraction. However, as
Eq. \eqref{eq:sf_frac_drude2} illustrates, the dependence of the superfluid
fraction on the excited states is more complex than just the energy gap.

%%%%%%%%%%%%%%%%%%%%%%%%%%%%%%%%%%%%%%%%%%%%%%%%%%%%%%%%%%%%%%%%%%
\section{Experimental observables}
\label{sec:exp}

The two quantities, condensate fraction and superfluid fraction, discussed
in the preceeding section describe two important and fundamental
properties of the system, however, they are not directly accessible to
simple experiments. In this section we discuss two observables which are
directly accessible to present day experiments and show their relation to
the fundamental quantities discussed above.

%%%%%%%%%%%%%%%%%%%%%%%%%%%%%%%%%%%%%%%%%%%%%%%%%%%%%%%%%%%%%%%%%%
\subsection{Matter-wave interference pattern}
\label{sec:intf}

The simplest experimental procedure to obtain information on the state of
the Bose gas in the lattice is to release the atoms from the lattice and
detect the matter-wave interference pattern after some time of flight
\cite{OrTu01, GrMa02a, GrMa02b}. The crucial question is, how much the
interference pattern can tell us about the superfluid or the condensed
portion of the system.

Neglecting interactions during the expansion of the gas we can describe 
the time evolution of the single-particle states, which are initially
localized at the different lattice sites, by Gaussian wave-packets. The
matter-wave intensity observed at a point $\yV$ after ballistic expansion
for a time $\tau$ can be written as \cite{RoBu03a}
\eq{ \label{eq:intf_intensity1}
  \IC(\yV) 
  = \matrixe{\Psi_0}{\AAO(\yV) \AO(\yV)}{\Psi_0} \;.
}
The amplitude operators are given by
\eq{
  \AO(\yV) 
  = \sum_{i=1}^{I} G_i(\yV,\tau)\; \aO_i \;,
}
where $G_i(\yV,\tau)$ is the amplitude of a Gaussian wave-packet that 
originates from site $i$ after free expansion for a time $\tau$.  We are
interested only in the generic structure of the intensity pattern, i.e.,
the presence or the absence of interference peaks. Therefore, we can
simplify the treatment significantly by dropping all the terms related to
the spatial envelope of the interference pattern. The Gaussian can thus be
replaced by a phase factor and the amplitude operator reduces to
\eq{
  \AO(\yV) 
  = \sum_{i=1}^{I} \ee^{\ii\, \phi_i(\yV,\tau)}\; \aO_i 
  = \sum_{i=1}^{I} \ee^{\ii\, \delta\phi(\yV,\tau)\, i }\; \aO_i \;,
}
Here $\phi_i(\yV,\tau)$ is the phase accumulated on the path from the
$i$th site to the observation point $\yV$. In the far-field approximation
this can be replaced by $\delta\phi(\yV,\tau)\, i$, where
$\delta\phi(\yV,\tau)$ is the phase difference between paths originating
from adjacent lattice sites. The matter-wave intensity
\eqref{eq:intf_intensity1} as function of $\delta\phi$ now reads
\eqmulti{ \label{eq:intf_intensity}
  \IC(\delta\phi)
  &= \frac{1}{I} \sum_{i,j=1}^{I} \ee^{\ii\,\delta\phi\, (i-j)}\;
    \matrixe{\Psi_0}{\aaO_j \aO_i}{\Psi_0} \\ 
  &= \frac{1}{I} \sum_{i,j=1}^{I} \ee^{\ii\,\delta\phi\, (i-j)}\;
    \rho^{(1)}_{ij} \;.
}
This result proves the direct relation between the interference pattern
and the quasimomentum distribution of the gas in the lattice given by Eq.
\eqref{eq:cond_quasimomdist} \cite{RoBu03a}. The occupation number
$\tilde{n}_q$ of the quasimomentum $q$ state is given by the intensity
$\IC(\delta\phi)$ for a phase difference $\delta\phi = qa$:
\eq{ \label{eq:intf_quasimomoccu}
  \tilde{n}_q = \IC(\delta\phi = qa) \;.
}
As a special case of this result, the condensate fraction, i.e. the
occupation of the $q=0$ state, is proportional to the intensity for
$\delta\phi=0$. Of course, this result agrees with the naive view that the
interference pattern simply gives the momentum distribution of the trapped
gas \cite{KaPr02}.

Furthermore Eq. \eqref{eq:intf_intensity} enables us to draw an important
conclusion about the relation between the matter-wave interference pattern
and the superfluid properties of the system. The interference pattern
depends exclusively on the one-body density matrix $\rho^{(1)}_{ij}$ for
the ground state. The superfluid fraction, in contrast, depends crucially
on the excited states of the system as we have discussed in detail in Sec.
\ref{sec:superfluid_3}. Thus the interference pattern cannot give complete
information on the superfluid properties of the system. It does not probe
the physics that is crucial for superfluidity. 

%%%%% figure %%%%%%%%%%%%%%%%%%%%%%%%%%%%%%%%%%%%%%%%%%%%%%%%%%%%%
\begin{figure}
\includegraphics[width=0.37\textwidth]{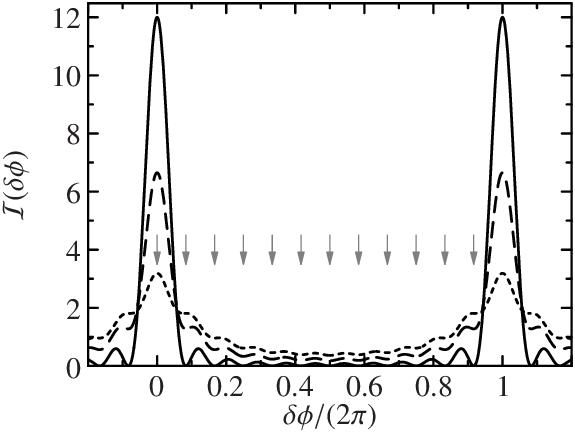}
\caption{Matter-wave intensity as function of the phase difference
$\delta\phi$ for a one-dimensional lattice system with $I=N=12$ and
different interaction strengths $V/J=0$ (solid line), $5$ (dashed), and 
$8$ (dotted).}
\label{fig:exp_intensity}
\end{figure}
%%%%%%%%%%%%%%%%%%%%%%%%%%%%%%%%%%%%%%%%%%%%%%%%%%%%%%%%%%%%%%%%%%

The matter-wave interference patterns which result from Eq.
\eqref{eq:intf_intensity} for three different interaction strength $V/J$
are depicted in Fig. \ref{fig:exp_intensity}. For the noninteracting system
(solid curve) it exhibits sharp interference peaks at $\delta\phi=0$ and
$2\pi$ which correspond to the principal peaks observed experimentally
\cite{GrMa02a}. Because we discarded the terms related to the envelope, the
intensity $\IC(\delta\phi)$ is simply a $2\pi$-periodic function in
$\delta\phi$. With increasing interaction strength $V/J$ (dashed and
dotted curves in Fig. \ref{fig:exp_intensity}) the amplitude of the
principal peaks is reduced and an ``incoherent background'' emerges.
Effectively this leads to a gradual broadening of the interference peaks
which was observed experimentally \cite{GrMa02a}. 

The emergence of the background has a straightforward interpretation in
terms of the quasimomentum distribution. The gray arrows in Fig.
\ref{fig:exp_intensity} mark the values of $\delta\phi$ which correspond
to the discrete quasimomenta $q$ allowed in the lattice. According to
\eqref{eq:intf_quasimomoccu} the intensity at these points just gives the
occupation numbers for the different quasimomentum states. Thus the
depletion of the principal peak with increasing $V/J$ just indicates the
depletion of the condensate. The emergence of a background shows that
states with nonvanishing quasimomenta are successively populated---as we
have already discussed in Sec. \ref{sec:quasimomdist}. In the limit of
strong interactions all quasimomentum states of the band are occupied
uniformly and the corresponding interference pattern is perfectly flat.

The fact that the vanishing of the superfluid fraction is not associated 
with the vanishing of the interference fringes is highlighted by these
numerical results. The values of $V/J$ used to compute the three intensity
distributions in Fig. \ref{fig:exp_intensity} correspond to superfluid
fractions of $f_{\SF}=1$ (solid curve), $f_{\SF}\approx0.5$ (dashed), and
$f_{\SF}\approx0$ (dotted). In the latter case, interference fringes are
still clearly visible although the system is a perfect insulator, i.e.,
the superfluid fraction is zero. Thus, we see that with increasing $V/J$
the superfluid component vanishes much earlier than the interference
fringes. In other words, the Mott insulator phase still exhibits a degree
of phase coherence.

As a simple quantity to characterize the interference pattern one might
introduce the visibility of the fringes. The standard definition of the
fringe visibility 
\eq{ \label{eq:intf_visibility}
  \VC = \frac{\IC_{\max} - \IC_{\min}}{\IC_{\max} + \IC_{\min}}
}
only relies on the maximum and minimum values of the intensity. For a
regular lattice with even number of sites we immediately find  $\IC_{\max}
= \IC(0)$ and $\IC_{\min} = \IC(\pi)$. On the basis of
\eqref{eq:intf_quasimomoccu} we can identify the maximum (minimum)
intensity with the largest (smallest) quasimomentum occupation number.
This definition of the visibility therefore provides a direct measure for 
the inhomogeneity of the quasimomentum distribution: $\VC$ is $1$ as long
as there are unoccupied quasimomentum states and it goes to zero if the
band is filled uniformly.

%%%%% figure %%%%%%%%%%%%%%%%%%%%%%%%%%%%%%%%%%%%%%%%%%%%%%%%%%%%%
\begin{figure}
\includegraphics[width=0.37\textwidth]{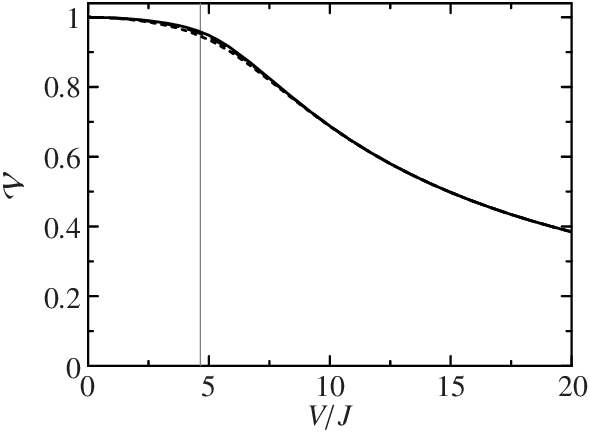}
\caption{Visibility $\VC$ of the interference fringes 
as function of $V/J$ for lattices with commensurate filling $N/I=1$ and
I=12 (solid), I=10 (dashed), and $I=8$ (dotted) lattice sites.}
\label{fig:intf_visibility}
\end{figure}
%%%%%%%%%%%%%%%%%%%%%%%%%%%%%%%%%%%%%%%%%%%%%%%%%%%%%%%%%%%%%%%%%%

The dependence of the visibility measure on the interaction strength is
depicted in Fig. \ref{fig:intf_visibility}. The visibility decreases only
very slowly with increasing $V/J$ and can have values up to
$\VC\approx0.7$ in the insulating phase where $f_{\SF}\approx0$. This
confirms the observation that superfluidity vanishes much faster than the
interference fringes and thus the coherence of the system \cite{RoBu03a}.
The fact that the curves in Fig. \ref{fig:intf_visibility} are practically
independent of the lattice size proves that this statement also holds for
large lattices and is not an artifact of the relatively limited size. With
regard to the quasimomentum distribution the slow decrease of
$\mathcal{V}$ indicates that very large ratios $V/J$ are required to
achieve a perfectly uniform occupation of the band. 

One can define other measures for the fringe visibility which are more
sensitive to particular changes in the interference patterns. However, the
most important and decisive measure for the coherence properties of the
system is the interference pattern itself.

%%%%%%%%%%%%%%%%%%%%%%%%%%%%%%%%%%%%%%%%%%%%%%%%%%%%%%%%%%%%%%%%%%
\subsection{Structure factor}
\label{sec:strucfac}

In the presence of a periodic modulation of the lattice potential the atom
density distribution will develop a spatial structure \cite{RoBu03b}.
Systems of this type will be discussed in detail in Sec. \ref{sec:2col}.
These periodic density modulations should be observable using Bragg
diffraction of light from the trapped atomic gas. Experiments of this kind
have already been performed using laser-cooled atomic gases in optical
lattices \cite{BiGa95,WeHe95}.

The quantity directly accessible through Bragg diffraction experiments is
the structure factor $S(k)$ which is given by \cite{OtWa95}
\eq{
  S(k) 
  = \frac{1}{I^2} \sum_{i,j=1}^{I} \ee^{\ii k (\xi_i-\xi_j)}
    \matrixe{\Psi_0}{\nO_i \nO_j}{\Psi_0} \;.
}
This quantity is sensitive to density-density correlations and is
therefore a suitable indicator for the presence of diagonal long-range
order. For a homogeneous occupation of all lattice sites, present in the
regular lattices discussed so far, the structure factor exhibits distinct
peaks at integer values of $ka/(2\pi)$.  The presence of any additional
periodic structure in the atomic density distribution will lead to the
appearance of new peaks at noninteger values of $ka/(2\pi)$. These signal
the presence of diagonal long-range order. From the position and the
relative amplitude of the additional peaks in the structure factor one can
derive detailed information on the spatial density distribution. We will
discuss this in detail in Sec. \ref{sec:2col_bg}.

%%%%%%%%%%%%%%%%%%%%%%%%%%%%%%%%%%%%%%%%%%%%%%%%%%%%%%%%%%%%%%%%%%
%%%%%%%%%%%%%%%%%%%%%%%%%%%%%%%%%%%%%%%%%%%%%%%%%%%%%%%%%%%%%%%%%%
%%%%%%%%%%%%%%%%%%%%%%%%%%%%%%%%%%%%%%%%%%%%%%%%%%%%%%%%%%%%%%%%%%
\section{Two-color superlattices}
\label{sec:2col}

Having discussed the relevant observables on a formal level and compared
their behavior for the superfluid to Mott-insulator transition in a
regular lattice we now want to explore another aspect of the phase diagram
of the zero temperature Bose gas. Optical lattices offer the unique
possibility of modifying the topology of the lattice in a flexible and
perfectly controlled manner.  They thus open new possibilities for studying
many-body physics in non-uniform lattice potentials. In contrast to most
solid-state systems, where the precise topology of the irregular lattice
is difficult to specify, the high degree of experimental control
facilitates investigations into specific aspects of ``disorder''-induced
phenomena.

%%%%%%%%%%%%%%%%%%%%%%%%%%%%%%%%%%%%%%%%%%%%%%%%%%%%%%%%%%%%%%%%%%
\subsection{Optical two-color lattices}

The simplest way to experimentally generate a non-uniform lattice
potential is through a superposition of two collinear optical standing
waves with different wavelengths---a so called two-color lattice. The
interference between these standing waves produces a superlattice with a
sinusoidal modulation of the depth of the lattice wells. More complex
topologies have already been realized experimentally by using a
superposition of several laser beams with different wave vectors
\cite{GuTr97}. Truly random lattices can be generated by superimposing a
speckle pattern \cite{DaZa03}. Thus optical lattices cover the whole range
from spatially modulated lattice potentials to disordered systems.

%%%%% figure %%%%%%%%%%%%%%%%%%%%%%%%%%%%%%%%%%%%%%%%%%%%%%%%%%%%%
\begin{figure}
\includegraphics[width=0.33\textwidth]{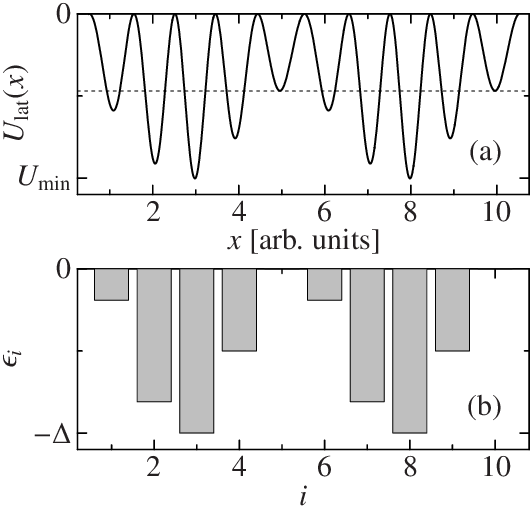}
\caption{(a) Lattice potential generated by superposition of two optical
standing waves (see text). (b) Corresponding distribution of 
the on-site energies $\epsilon_i$ with modulation amplitude $\Delta$ (after
subtraction of an over-all constant).}
\label{fig:2col_topology}
\end{figure}
%%%%%%%%%%%%%%%%%%%%%%%%%%%%%%%%%%%%%%%%%%%%%%%%%%%%%%%%%%%%%%%%%%

In the following we consider a superlattice generated by the interference
of two standing waves with wavelengths $\lambda_1$ and $\lambda_2 =
\frac{5}{7} \lambda_1$. The intensity of the second laser field is assumed
to be much smaller than the intensity of the first laser, in this example
$I_2 = 0.04 I_1$. This combination results in a superlattice consisting of
identical unit cells each composed of $5$ elementary lattice sites. Figure
\ref{fig:2col_topology}(a) depicts the lattice potential
$U_{\text{lat}}(x)$ for a certain value of the relative phase. 

In the language of the Bose-Hubbard model this spatial modulation of the
well depths corresponds to a modulation of the on-site energies
$\epsilon_i$. We assume a sinusoidal variation of the $\epsilon_i$ in the
range $[-\Delta,0]$ as depicted in Fig. \ref{fig:2col_topology}(b). The
parameter $\Delta$ characterizes the amplitude of the modulation, for
$\Delta=0$ we recover the regular lattice potential.  The modulation of
the lattice potential also influences the tunneling coefficient $J$ which,
in general, varies for the different lattice sites. For the lattices
considered here the influence of this variation can be neglected in good
approximation \cite{DaZa03}.

%%%%%%%%%%%%%%%%%%%%%%%%%%%%%%%%%%%%%%%%%%%%%%%%%%%%%%%%%%%%%%%%%%
\subsection{Localized phase}
\label{sec:2col_loc}

\begin{figure}
\includegraphics[width=0.34\textwidth]{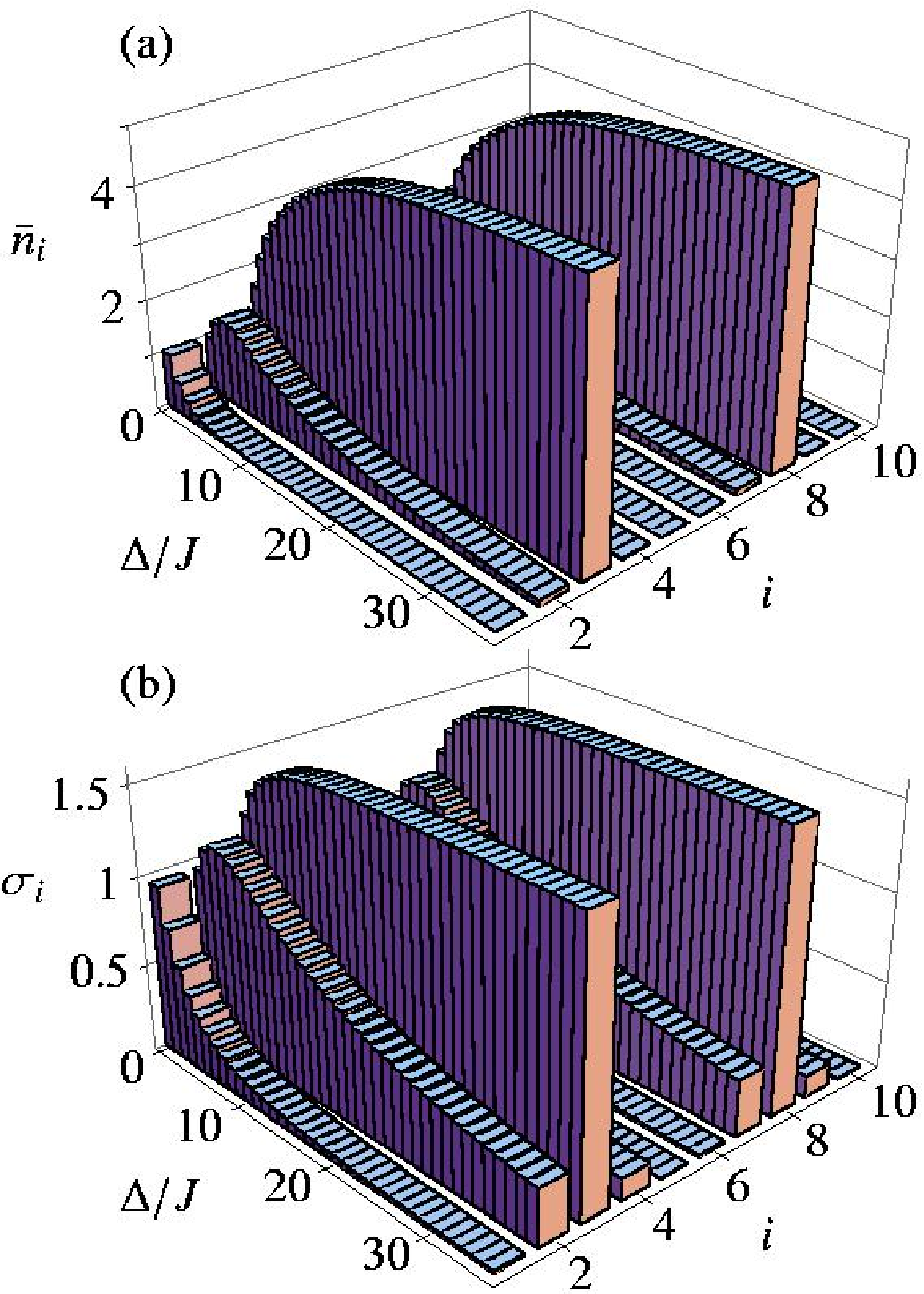}
\caption{Mean occupation numbers $\bar{n}_i$ (a) and 
number fluctuations $\sigma_i$ (b) for the different lattice sites
$i=1,...,10$ as function of the amplitude $\Delta/J$ of the sinusoidal
modulation for the noninteracting gas.}
\label{fig:loc_nMeanFluc}
\end{figure}

As a first step we consider a noninteracting Bose gas in the superlattice 
depicted in Fig. \ref{fig:2col_topology}. We solve the eigenvalue problem
for the  Bose-Hubbard Hamiltonian including two unit cells with filling
factor $1$, i.e., $I=N=10$. For $V/J=0$ we successively increase the
amplitude $\Delta$ of the spatial modulation of the on-site energies. The
resulting distributions of mean occupation numbers $\bar{n}_i$ and number
fluctuations $\sigma_i$ for the individual sites are depicted in Fig.
\ref{fig:loc_nMeanFluc}. For a regular system, without any modulation of
the $\epsilon_i$, the lattice is occupied homogeneously. With increasing
modulation amplitude $\Delta/J$ the particles preferentially occupy those
lattice sites with lower $\epsilon_i$ as one would classically expect. The
degree of localization is governed by the competition between the on-site
energies and the tunneling term of the Bose-Hubbard Hamiltonian
\eqref{eq:hamiltonian}. For sufficiently large $\Delta/J$ only the site
with the lowest $\epsilon_i$ in each unit cell is populated.  The number
fluctuations $\sigma_i$ behave accordingly: with increasing $\Delta/J$ the
number fluctuations at the deepest lattice well grow whereas the
fluctuations at all other wells slowly decrease. In the limit of large
$\Delta/J$ the ground state is a superposition of all Fock states which
have nonzero occupation numbers only at the deepest well of each unit
cell.

%%%%% figure %%%%%%%%%%%%%%%%%%%%%%%%%%%%%%%%%%%%%%%%%%%%%%%%%%%%%
%%%%%%%%%%%%%%%%%%%%%%%%%%%%%%%%%%%%%%%%%%%%%%%%%%%%%%%%%%%%%%%%%%

\begin{figure}
\includegraphics[width=0.35\textwidth]{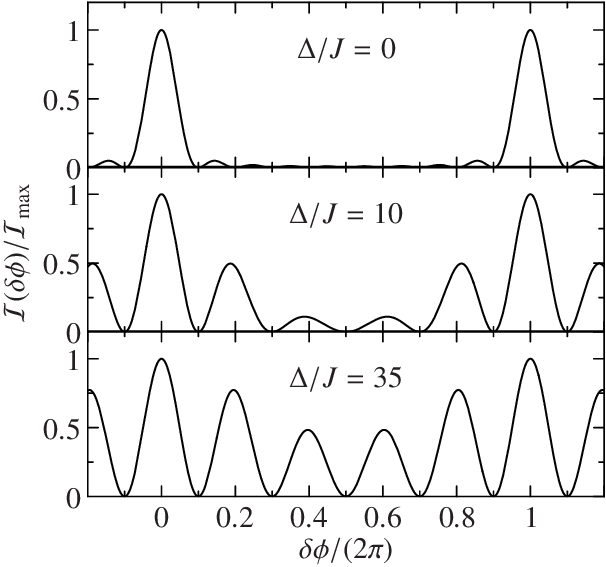}
\caption{Interference pattern after release from the lattice for different
values of the amplitude $\Delta/J$ of the sinusoidal modulation of the
on-site energies.}
\label{fig:loc_intensity}
\end{figure}

There are several possibilities for detecting this localized phase in an
experiment. The simplest signature appears in the interference pattern
after the expansion of the gas. Figure \ref{fig:loc_intensity} shows the
intensity distribution that results from Eq. \eqref{eq:intf_intensity} 
for three different values of $\Delta/J$. In the absence of the spatial
modulation (upper panel) we observe the interference pattern of a regular
lattice discussed in Sec. \ref{sec:intf}. With increasing $\Delta/J$, that
is increasing degree of localization, additional peaks emerge. First the
peaks next to initial principal peaks appear, then the next-order peaks
emerge.  In the limit of perfect localization we obtain a regular
interference pattern with a peak separation corresponding to the inverse
of the number of elementary sites in a unit cell. In this manner the
relative strength of the emerging peaks provides a direct experimental
measure for the degree of localization in these superlattices.

%%%%% figure %%%%%%%%%%%%%%%%%%%%%%%%%%%%%%%%%%%%%%%%%%%%%%%%%%%%%
%%%%%%%%%%%%%%%%%%%%%%%%%%%%%%%%%%%%%%%%%%%%%%%%%%%%%%%%%%%%%%%%%%

Another experimental observable that is sensitive to the degree of
localization is the structure factor $S(ka)$ introduced in Sec.
\ref{sec:strucfac}. We will come back to this quantity in the next section.

%%%%%%%%%%%%%%%%%%%%%%%%%%%%%%%%%%%%%%%%%%%%%%%%%%%%%%%%%%%%%%%%%%
\subsection{Quasi Bose-glass phase}
\label{sec:2col_bg}

It is intuitively clear that repulsive interactions will have a strong
influence on the localized phase: A concentration of many particles at a
single lattice site will generate a large interaction energy contribution.
Thus the interaction enforces delocalization of the particles.

\begin{figure}
\includegraphics[width=0.34\textwidth]{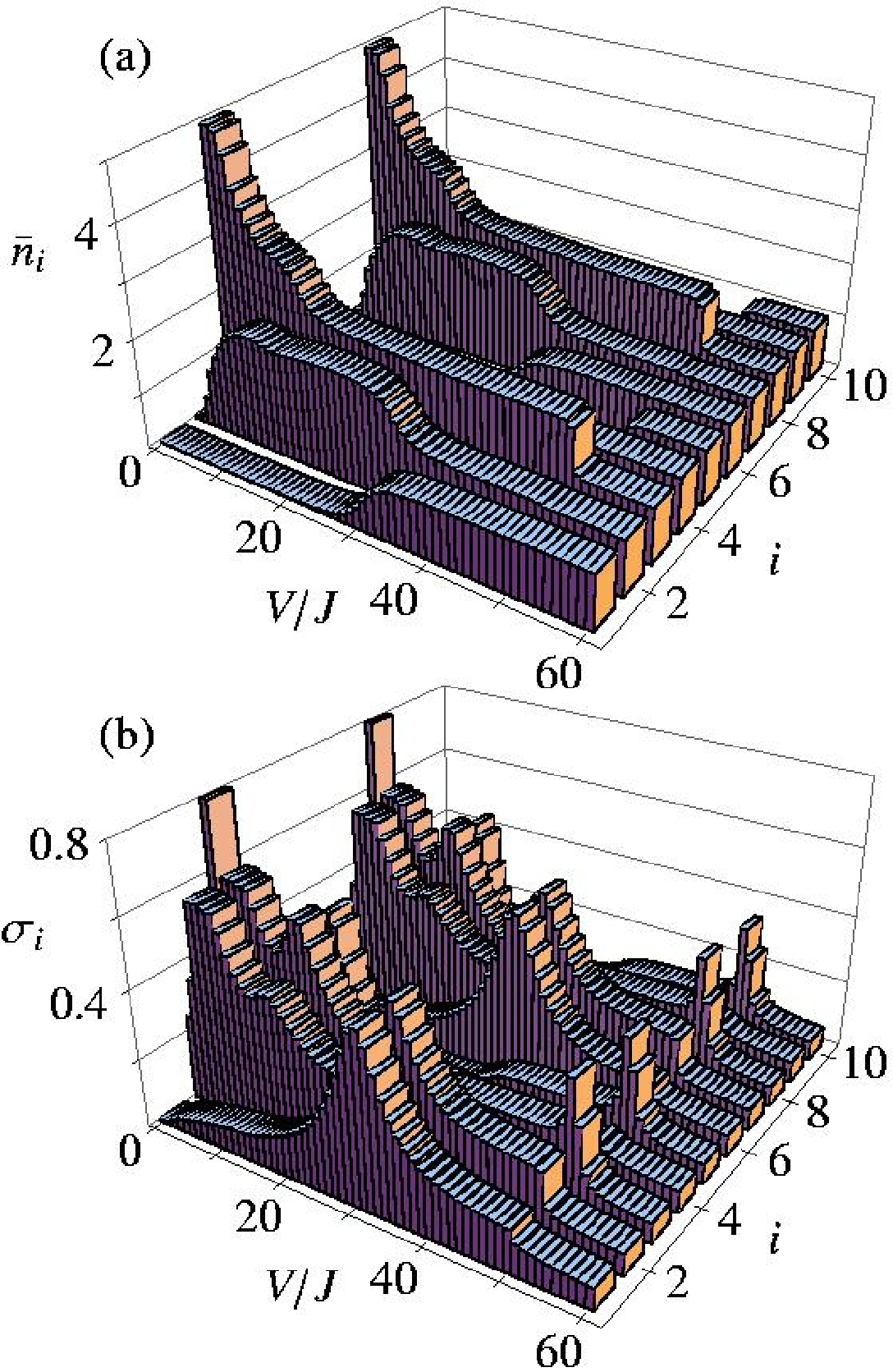}
\caption{Mean occupation numbers $\bar{n}_i$ (a) and the
number fluctuations $\sigma_i$ (b) for the individual lattice sites
$i=1,...,10$ as function of  the interaction strength $V/J$ for fixed
modulation amplitude $\Delta/J=50$.}
\label{fig:bg_nMeanFluc}
\end{figure}

To illustrate the competition between lattice modulation and repulsive
two-body interaction Fig. \ref{fig:bg_nMeanFluc} depicts the  change of
the mean occupation numbers $\bar{n}_i$ and the number fluctuations
$\sigma_i$ with increasing interaction strength $V/J$ for a fixed
modulation amplitude $\Delta/J=50$. 

The perfectly localized configuration persists only at very small
interaction strengths. Already for $V/J \gtrsim 2$ it is energetically 
favorable to reduce the maximum occupation number and thus the interaction
energy by redistributing particles to sites with higher on-site energies
$\epsilon_i$. This region of interaction-induced delocalization is
sometimes termed Anderson-glass phase \cite{ScBa91,DaZa03}. Initially the
changes in the mean occupation numbers happen continuously. However, for
$V/J\gtrsim10$ this behavior changes. As shown in Fig.
\ref{fig:bg_nMeanFluc}(a) there are extended intervals of interaction
strengths where the mean occupation numbers are constant and approximately
integer. Between these regions of ``stable'' configurations there are narrow
windows in which some of the $\bar{n}_i$ change by roughly $\pm1$. In
the present example one of these almost steplike rearrangements happens at
$V/J\approx30$. This behavior of the mean occupation numbers is also
reflected in the number fluctuations depicted in Fig.
\ref{fig:bg_nMeanFluc}(b). Within the stable regions all $\sigma_i$ are
rather small, i.e., the dominant contribution to the expansion
\eqref{eq:state} of the ground state comes from a single Fock state. The
steplike rearrangements are accompanied by an increase of the number
fluctuations of those sites that change their mean occupation number. 

%%%%% figure %%%%%%%%%%%%%%%%%%%%%%%%%%%%%%%%%%%%%%%%%%%%%%%%%%%%%
%%%%%%%%%%%%%%%%%%%%%%%%%%%%%%%%%%%%%%%%%%%%%%%%%%%%%%%%%%%%%%%%%%

It is the subtle interplay between interaction and lattice irregularity
that drives these successive rearrangements between configurations with
integer occupation numbers. The sequence of rearrangements depends
crucially on the topology of the lattice, i.e., the set of on-site
energies. If the distribution of the $\epsilon_i$ becomes more homogeneous
(meaning that the number of different values of the $\epsilon_i$ increases) 
then the number of possible rearrangements increases and the 
stable regions shrink. In the limit of a lattice with random disorder this
leads to the so-called Bose-glass phase \cite{ScBa91,KrTr91}. For the
superlattices with a rather small number of different on-site energies we
will use the term quasi Bose-glass phase. The study of the superfluid
fraction in Sec. \ref{sec:2col_phasediag} shows that the quasi Bose-glass
phase is insulating and that the energy gap is small.

If the interaction strength reaches a value $V = \Delta$ a final and very
sharp rearrangement happens and the system assumes a homogeneous
occupation of all sites. This is the transition to the Mott-insulator 
phase where despite of the presence of the irregular lattice the ground
state is dominated by the Fock state with uniform occupation numbers
$n_i=N/I$. This transition between quasi Bose-glass and Mott-insulator
phase happens independently of the lattice topology if the interaction
strength $V$ exceeds the amplitude $\Delta$ of the lattice modulation.

%%%%% figure %%%%%%%%%%%%%%%%%%%%%%%%%%%%%%%%%%%%%%%%%%%%%%%%%%%%%
\begin{figure}
\includegraphics[width=0.35\textwidth]{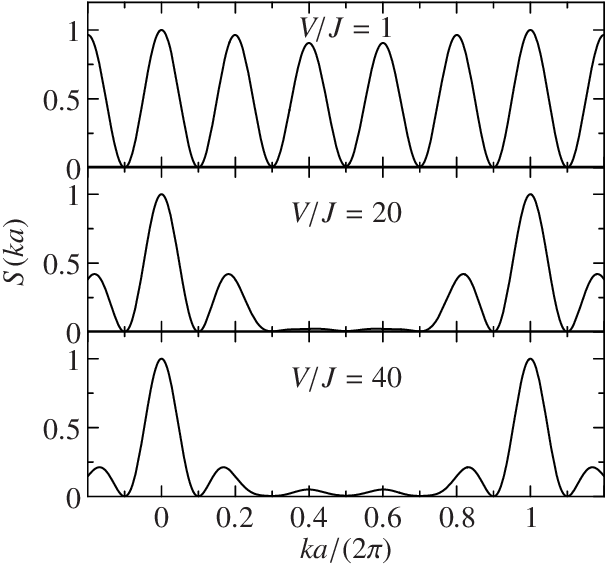}
\caption{Structure factor $S(ka)$ for different
values of the interaction strength $V/J$ and fixed amplitude $\Delta/J=50$.}
\label{fig:bg_strucfac}
\end{figure}
%%%%%%%%%%%%%%%%%%%%%%%%%%%%%%%%%%%%%%%%%%%%%%%%%%%%%%%%%%%%%%%%%%

The sequence of rearrangements in the quasi Bose-glass phase can be
directly observed in experiment. The observable which is most sensitive to
the changes in the spatial distribution of the atoms in the superlattice
is the structure factor $S(ka)$ introduced in Sec. \ref{sec:strucfac}. For
a state with a homogeneous population of the lattice sites, e.g., in the
Mott-insulator phase, $S(ka)$ shows peaks at integer values of $ka/(2\pi)$
only. If mean-occupation numbers, i.e. the diagonal elements of the
one-body density matrix, exhibit some periodic long-range structure  then
new peaks will appear at noninteger values of $ka/(2\pi)$ from which
information on the long-range order can be extracted.

The localized phase in a superlattice at small interaction strengths is
the most pronounced realization of diagonal long-range order: For the
superlattice discussed here each $5$th lattice site is occupied and all
others are practically empty. The corresponding structure factor $S(ka)$ is
shown in the upper panel of Fig. \ref{fig:bg_strucfac}. Peaks of almost
equal height emerge at $ka/(2\pi) = \nu/5$ ($\nu$ integer) indicating the
perfect localization at a single site in each unit cell. With increasing
interaction strength $V/J$ the particles are gradually redistributed to
other lattice sites. This leads to a suppression of the peaks at
non-integer values of $ka/(2\pi)$ as the two lower panels Fig.
\ref{fig:bg_strucfac} illustrate. Thus the amplitude of the peaks in
$S(ka)$ provides detailed information on the complicated spatial structure
within the quasi Bose-glass phase.

%%%%% figure %%%%%%%%%%%%%%%%%%%%%%%%%%%%%%%%%%%%%%%%%%%%%%%%%%%%%
\begin{figure}
\includegraphics[width=0.37\textwidth]{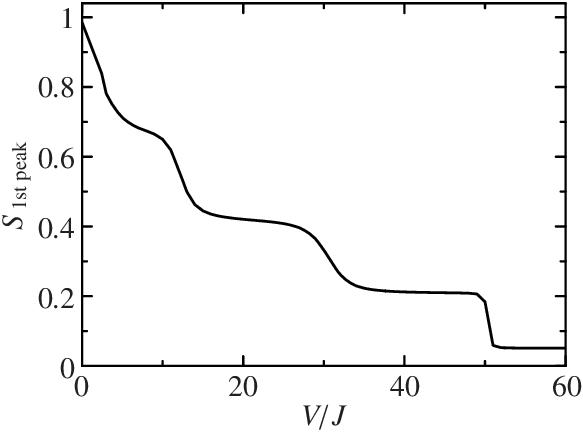}
\caption{Amplitude of the first peak of the structure factor $S(ka)$ 
located around $ka \approx 2\pi/5$ as function of the interaction strength
$V/J$ for fixed $\Delta/J=50$.}
\label{fig:bg_strucpeak}
\end{figure}
%%%%%%%%%%%%%%%%%%%%%%%%%%%%%%%%%%%%%%%%%%%%%%%%%%%%%%%%%%%%%%%%%%

To highlight this point further, Fig. \ref{fig:bg_strucpeak} shows the
dependence of the amplitude of the first peak of $S(ka)$ located around
$ka/(2\pi)\approx 1/5$ as function of $V/J$ for fixed amplitude $\Delta/J$
of the lattice modulation. The parameters correspond to those used in Fig.
\ref{fig:bg_nMeanFluc} and cover the full range from the localized phase
to the Mott-insulator. Obviously each of the rearrangements of the mean
occupation numbers within the quasi Bose-glass phase leaves a distinct
signature in the peak amplitude. The regions of stable configurations as
well as the steplike rearrangements that are visible in Fig.
\ref{fig:bg_nMeanFluc}(a) directly show up in $S(ka)$. Thus the
structure factor, which is experimentally accessible through Bragg
scattering of light from the bosons in the lattice, provides a
comprehensive insight into the spatial density structure and its complex
dependency on the interaction strength and the modulation amplitude.

%%%%%%%%%%%%%%%%%%%%%%%%%%%%%%%%%%%%%%%%%%%%%%%%%%%%%%%%%%%%%%%%%%
\subsection{Phase diagrams in the $V$-$\Delta$ plane}
\label{sec:2col_phasediag} 

%%%%% figure %%%%%%%%%%%%%%%%%%%%%%%%%%%%%%%%%%%%%%%%%%%%%%%%%%%%%
\begin{figure}
\includegraphics[width=0.41\textwidth]{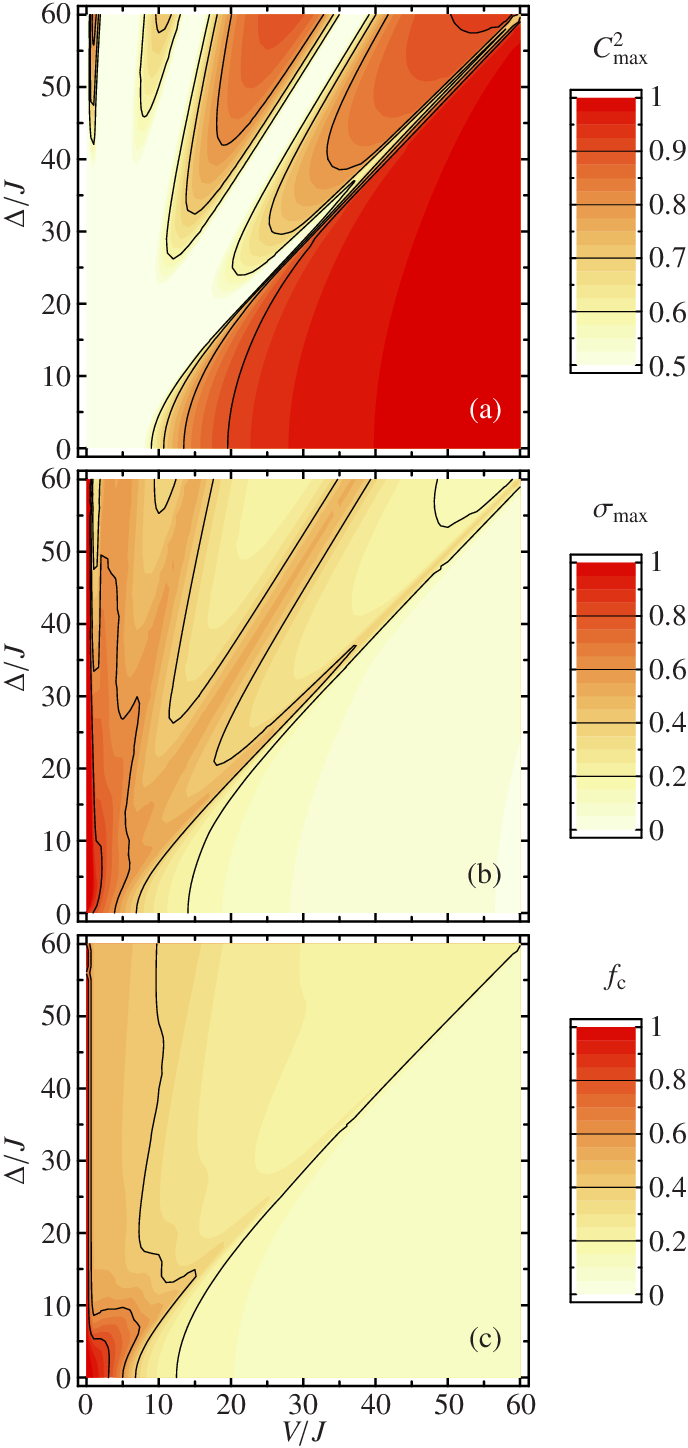}
\caption{Contour plots of the largest coefficient $C_{\max}$ (a), the
maximum number fluctuations $\sigma_{\max}$ (b), and the condensate
fraction $f_{\cond}$ (c) in the $V$-$\Delta$ plane for the sinusoidal
lattice with $I=N=10$.}
\label{fig:phasediag_1}
\end{figure}
%%%%%%%%%%%%%%%%%%%%%%%%%%%%%%%%%%%%%%%%%%%%%%%%%%%%%%%%%%%%%%%%%%

We shall now illustrate the behavior of the superfluid fraction,
condensate fraction, and energy gap using two-dimensional phase diagrams
in the $V$-$\Delta$ plane. This representation exposes the complicated
interplay between lattice modulation and two-body interaction which leads
to the rich phase structure.

The gross phase structure is already revealed in the behavior of the
largest coefficient $C_{\max}^2$ in the expansion \eqref{eq:state} of the
ground state which is depicted in Fig. \ref{fig:phasediag_1}(a). The dark
shadings correspond to values of  $C_{\max}^2$ close to $1$, i.e. ground
states which are almost pure Fock states. The prominent region of large
$C_{\max}^2$ for values $V>\Delta$ corresponds to the Mott-insulator
phase. The lobe-like structures at large modulation amplitudes $\Delta$
belong to the quasi Bose-glass phase. They correspond to the regions of
stable configurations  discussed in connection with Fig.
\ref{fig:bg_nMeanFluc}  in which the mean occupation numbers are constant
and approximately integral. The valleys of small $C_{\max}^2$ which
separate the lobes are connected to the rearrangements between different
stable configurations. As mentioned earlier the number and position of
these lobes depends on the topology of the lattice potential.

In close connection with the behavior of $C_{\max}^2$ are the maximum
number fluctuations $\sigma_{\max} = \max(\sigma_i)$ depicted in Fig.
\ref{fig:phasediag_1}(b). Values of $C_{\max}^2$ close to $1$ imply that
the number fluctuations at all lattice sites are small. Thus the
Mott-insulator phase and the lobes of the quasi Bose-glass phase appear as
regions of small number fluctuations (light shadings). In contrast, the
rearrangement valleys in the quasi Bose-glass phase, which are associated
with small $C_{\max}^2$, show increased number fluctuations. Number
fluctuations of the order $1$ emerge only within a narrow band at small
values of $V/J$. For large values of $\Delta/J$ this corresponds to the
localized phase discussed in Sec. \ref{sec:2col_loc}.

The behavior of the condensate fraction $f_{\cond}$ is shown in Fig.
\ref{fig:phasediag_1}(c). In line with variations of the maximum number
fluctuations, $f_{\cond}$ assumes very small values in the Mott-insulator
phase. Within the quasi Bose-glass it decreases monotonically with
increasing $V/J$, i.e., it does not show the lobe structure as the number
fluctuations. Values of the condensate fraction above $0.6$ result only in
a small area at small $V/J$ and $\Delta/J$ and a very narrow stripe at
$V/J\lesssim2$. Thus only the localized phase and the expected superfluid
phase exhibit sizeable condensate fractions.

%%%%% figure %%%%%%%%%%%%%%%%%%%%%%%%%%%%%%%%%%%%%%%%%%%%%%%%%%%%%
\begin{figure}
\includegraphics[width=0.41\textwidth]{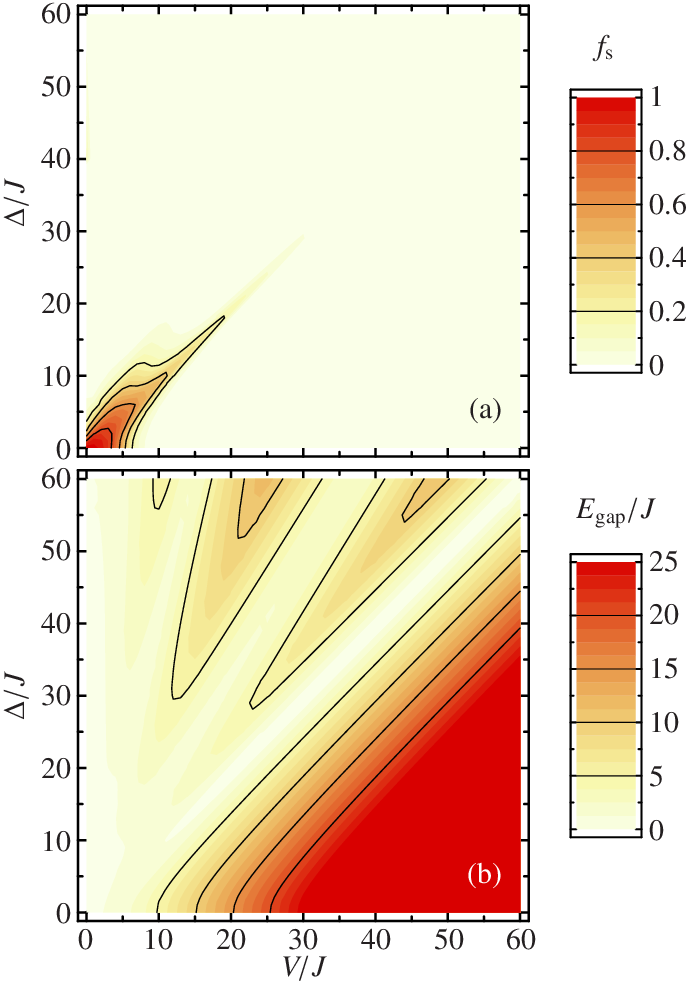}
\caption{Contour plots of the superfluid fraction $f_{\SF}$ (a) and the 
energy gap $E_{\text{gap}}$ (b) for the sinusoidal
lattice with $I=N=10$.}
\label{fig:phasediag_2}
\end{figure}
%%%%%%%%%%%%%%%%%%%%%%%%%%%%%%%%%%%%%%%%%%%%%%%%%%%%%%%%%%%%%%%%%%

So far, these quantities are determined by the properties of the ground
state of Bose gas. Following the discussion in Sec. \ref{sec:sf} they
therefore do not fully determine the superfluid properties of the system.
Figure \ref{fig:phasediag_2} shows the contour plots for two quantities
which probe the excitation spectrum, these are the superfluid fraction and
the energy gap. The plot for superfluid fraction $f_{\SF}$ in Fig.
\ref{fig:phasediag_2}(a) pinpoints the boundaries of the superfluid phase.
Large values of the superfluid fraction (dark shadings) result only in a
restricted parameter region at small values of $V$ and $\Delta$.
Everywhere else---that is, in the Mott-insulator, the quasi Bose-glass,
and the localized phase---the system is a perfect insulator. In comparing
the boundaries of the superfluid phase defined in Fig.
\ref{fig:phasediag_2}(a) with the structure of the other observables, like
number fluctuations and condensate fraction in Fig. \ref{fig:phasediag_1},
it becomes evident that none of the ground state quantities can identify
the superfluid region. Not only that the magnitude of the superfluid
fraction cannot be determined, the whole structure of the plots is
different.

This also applies to the energy gap $E_{\text{gap}}$ plotted in Fig. 
\ref{fig:phasediag_2}(b). Although the opening of a gap between the ground
state and the first excited state is an indicator for the superfluid to
Mott-insulator transition in the uniform lattice, there is no such
correspondence for the irregular lattice. It is only the Mott-insulator
phase that exhibits a large energy gap. The localized and the quasi
Bose-glass phase, which are also insulating, have small energy gaps just
like the superfluid phase.  

The energy gap of the quasi Bose-glass phase deserves closer inspection.
Figure \ref{fig:phasediag_2}(b) reveals that there is a small but finite
energy gap within the stable lobes of the quasi Bose-glass phase. The size
of the gaps, as well as the position and number of the lobes, depend on
the differences between the various on-site energies. If the size of the
unit cells is increased or if one goes to large random lattices then the
differences between the $\epsilon_i$ become smaller, the number of  lobes
increases, their size decreases, and the energy gap vanishes. Thus there
is a continuous transition from the quasi Bose-glass to the strict
Bose-glass phase with vanishing $E_{\text{gap}}$. 

Another interesting observation results from the comparison of the
superfluid fraction in Fig. \ref{fig:phasediag_2}(a) with the condensate
fraction in Fig. \ref{fig:phasediag_1}(c). In an irregular lattices with
weak interactions there is a region where the condensate fraction is larger
than the superfluid fraction. An extreme case is the localized phase at
$V/J\lesssim2$, where $f_{\SF}\approx0$ and $f_{\cond}\approx 1$. This
means that the localized bosons in the different unit cells are still
phase coherent but the superfluid flow is inhibited by the lattice
topology. A similar phenomenon was recently observed in a Monte Carlo
study of a continuous Bose gas with random impurities \cite{AsBo02} which act
like the irregular lattice potential. 

So far we have only considered the phase diagrams for a filling factor
$N/I=1$. The qualitative structure of the phase diagram does not change if
we go to larger integer filling factors. Quantitatively the superfluid
region expands towards larger values of the interaction strength $V/J$ and
the modulation amplitude $\Delta/J$ if the filling factor is increased.
For a filling factor $N/I=4$, for example, the contour corresponding to a
superfluid fraction $f_{\SF}=0.6$ extends up to $V/J\approx15$ and
$\Delta/J\approx40$. The insulating phases are shifted accordingly.

%%%%%%%%%%%%%%%%%%%%%%%%%%%%%%%%%%%%%%%%%%%%%%%%%%%%%%%%%%%%%%%%%%
\subsection{Non-commensurate filling factors}

%%%%% figure %%%%%%%%%%%%%%%%%%%%%%%%%%%%%%%%%%%%%%%%%%%%%%%%%%%%%
\begin{figure}
\includegraphics[width=0.41\textwidth]{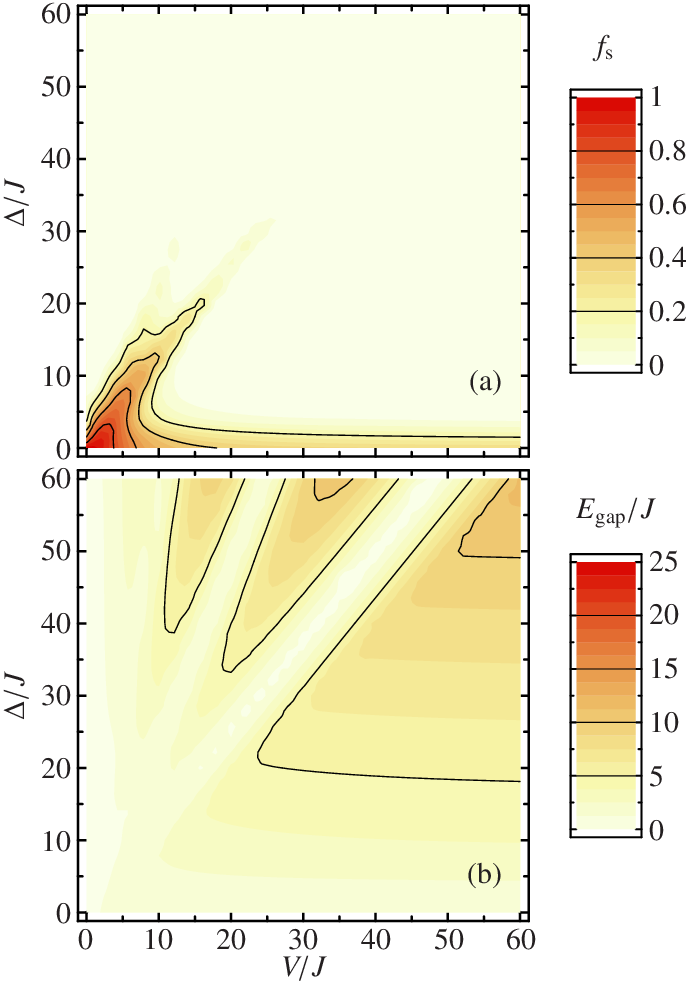}
\caption{Contour plots of the superfluid fraction $f_{\SF}$ (a) and the 
energy gap $E_{\text{gap}}$ (b) in the $V$-$\Delta$ plane for the
sinusoidal lattice with $I=10$ and non-commensurable particle number
$N=12$.}
\label{fig:phasediag_noncom}
\end{figure}
%%%%%%%%%%%%%%%%%%%%%%%%%%%%%%%%%%%%%%%%%%%%%%%%%%%%%%%%%%%%%%%%%%

It is well-known that the Mott-insulator phase in a uniform lattice exists
only for commensurate filling factors. If one adds a few particles on
top of a commensurate configuration then these particles can be moved to any
lattice site without changing the energy---even in the presence of strong
repulsive interactions. In this way the additional particles gives rise to
a finite superfluid fraction in the limit of large $V/J$. 

This intuitive picture is confirmed by the exact calculations for the
sinusoidal superlattice with $I=10$ sites and $N=12$
particles. Contour plots for the superfluid fraction and the energy gap in
the $V$-$\Delta$  plane are presented in Fig. \ref{fig:phasediag_noncom}.
For the uniform lattice ($\Delta/J=0$) the superfluid fraction shown in
panel (a) remains finite even at large $V/J$.

If one starts to increase the amplitude $\Delta$ of the sinusoidal
modulation then the superfluid fraction decreases and eventually vanishes
above $\Delta/J\gtrsim 5$. In this regime the growing lattice
irregularity causes a pinning of the additional particles to the lattice
sites with minimal on-site energies $\epsilon_i$. The ground state has an
increased occupation number at the deepest lattice well of each unit cell. 

The behavior of the energy gap shown in Fig. \ref{fig:phasediag_noncom}(b)
confirms this picture. For strong repulsive interactions the energy gap
vanishes at small amplitudes $\Delta/J$ in stark contrast to the
commensurate system. With increasing amplitude of the sinusoidal
modulation the energy gap starts to grow linearly with $\Delta/J$. In this
simple case $E_{\text{gap}}$ corresponds to the energy required to remove
the additional particle from the lattice site it is affixed to. The slope
of the increase is proportional to the energy difference between the two
lowest $\epsilon_i$. This behavior is equivalent to the structure of the
lobes of the quasi Bose-glass phase.

Thus, in general, the change from a commensurate to a non-commensurate
filling factor causes a vanishing of the Mott-insulator phase. It is
replaced by a superfluid phase at small $\Delta/J$, where the additional
particles constitute a superfluid on top of an Mott-insulating layer. At
larger $\Delta/J$ the additional particles are pinned to the deepest
lattice wells, superfluidity ceases, and an additional lobe of the quasi
Bose-glass phase emerges. All other regions of the phase diagram are 
qualitatively unaffected by the change from a commensurate to a
non-commensurate filling factor.

%%%%%%%%%%%%%%%%%%%%%%%%%%%%%%%%%%%%%%%%%%%%%%%%%%%%%%%%%%%%%%%%%%
%%%%%%%%%%%%%%%%%%%%%%%%%%%%%%%%%%%%%%%%%%%%%%%%%%%%%%%%%%%%%%%%%%
%%%%%%%%%%%%%%%%%%%%%%%%%%%%%%%%%%%%%%%%%%%%%%%%%%%%%%%%%%%%%%%%%%
\section{Conclusions}

We have shown that the superfluid fraction---the natural order parameter
for superfluid to insulator phase transitions---is determined by the
response of the system to an external perturbation, i.e., a twist in the
boundary conditions, or equivalently an additional phase factor, in the
tunneling term of the Bose-Hubbard Hamiltonian. This means that the
superfluid fraction necessarily depends on the full excitation spectrum of
the lattice system. The formal manifestation of this fact is given in Eq.
\eqref{eq:sf_frac_drude}, where the superfluid fraction is expressed as a
sum of a first order term depending only of the ground state and a second
order term which involves all excited states. The exact numerical solution
of the Bose-Hubbard model for uniform one-dimensional lattices reveals
that the second order term is crucial for the vanishing of the superfluid
fraction in the Mott-insulator phase. The first order contribution gives
only an upper bound for the superfluid fraction and can have a rather
large value in the insulating phase. 

This allows important conclusions for the prospects of measuring the
superfluid fraction and the details of superfluid to insulator phase
transitions. Present experimental observables are sensitive only to the
ground state and therefore cannot probe the physics that governs the
superfluid properties. The matter-wave interference pattern, for example,
provides direct information on the quasimomentum distribution of the
ground state but it does not measure the superfluid properties.
In the case of the superfluid to Mott-insulator transition our numerical
results reveal that the superfluid fraction vanishes much earlier than the
interference fringes, i.e., an interference pattern is still visible in the
Mott-insulator phase. This demonstrates that one has to clearly distinguish
between superfluidity and coherence properties. 

In the second part of the paper we have employed these tools to explore
the zero-temperature phase diagram of Bose gases in non-uniform lattice
potentials. We have demonstrated that even a simple superlattice potential
which results from the superposition of two standing-wave lattices with
different wavelengths gives rise to a very rich phase diagram. As function
of the interaction strengths $V/J$ and the amplitude $\Delta/J$ of the
spatial variation of the on-site energies two additional insulating phases
can be identified: a localized phase and a quasi Bose-glass phase. All
insulating phases can be clearly distinguished through their
characteristic signatures in matter-wave interference experiments or
through the structure factor determined by Bragg diffraction of light.

Our results further support the view that Bose gases in optical lattices
are a versatile tool for studying quantum mechanical many-body phenomena
in strongly correlated systems. They allow us to address fundamental
questions such as the connection between superfluidity and Bose-Einstein
condensation. Moreover, they enable controlled studies of disorder-induced
phenomena which result from a subtle competition between kinetic energy,
two-body interaction, and lattice topology. To this end two-color lattices
with relatively small unit cells seem a promising starting point because
they facilitate a clear experimental distinction between the various
phases.

%%%%%%%%%%%%%%%%%%%%%%%%%%%%%%%%%%%%%%%%%%%%%%%%%%%%%%%%%%%%%%%%%%
%%%%%%%%%%%%%%%%%%%%%%%%%%%%%%%%%%%%%%%%%%%%%%%%%%%%%%%%%%%%%%%%%%
%%%%%%%%%%%%%%%%%%%%%%%%%%%%%%%%%%%%%%%%%%%%%%%%%%%%%%%%%%%%%%%%%%
\section*{Acknowledgments}

This work was supported by the DFG, the UK EPSRC, and the EU via the
``Cold Quantum Gases'' network. Keith Burnett thanks the Royal Society and
Wolfson Foundation for support. We acknowledge useful discussions with
S.A. Gardiner, J.A. Dunningham, T. K\"ohler, T. Gasenzer, and K.
Braun-Munzinger.

%%%%%%%%%%%%%%%%%%%%%%%%%%%%%%%%%%%%%%%%%%%%%%%%%%%%%%%%%%%%%%%%%%
%%%%%%%%%%%%%%%%%%%%%%%%%%%%%%%%%%%%%%%%%%%%%%%%%%%%%%%%%%%%%%%%%%
%%%%%%%%%%%%%%%%%%%%%%%%%%%%%%%%%%%%%%%%%%%%%%%%%%%%%%%%%%%%%%%%%%

%%%%%%%%%%%%%%%%%%%%%%%%%%%%%%%%%%%%%%%%%%%%%%%%%%%%%%%%%%%%%%%%%%
%%%%%%%%%%%%%%%%%%%%%%%%%%%%%%%%%%%%%%%%%%%%%%%%%%%%%%%%%%%%%%%%%%
%%%%%%%%%%%%%%%%%%%%%%%%%%%%%%%%%%%%%%%%%%%%%%%%%%%%%%%%%%%%%%%%%%
\end{document}